\documentclass[a4paper,fleqn,usenatbib]{mnras}
\usepackage{newtxtext,newtxmath}
\usepackage[T1]{fontenc}
\usepackage{ae,aecompl}

\usepackage{graphicx}	% Including figure files
\usepackage{amsmath}	% Advanced maths commands
\usepackage{amssymb}	% Extra maths symbols
\usepackage{threeparttable}
\usepackage{longtable}
\usepackage{threeparttablex}  % for "ThreePartTable" environment
\usepackage{dcolumn}
\usepackage{multirow}
\usepackage{supertabular, booktabs}
\usepackage{tabularx}
\usepackage{caption}
\usepackage{subcaption}
\usepackage{multicol}
\usepackage{pdflscape}

\newcolumntype{C}[1]{>{\hsize=#1\hsize\centering\arraybackslash}X}%
\newcolumntype{L}[1]{>{\hsize=#1\hsize\raggedright\arraybackslash}X}%
\newcolumntype{R}[1]{>{\hsize=#1\hsize\raggedleft\arraybackslash}X}%
\newcommand{\kms}{km s$^{-1}$}
\newcommand{\lxlbol}{log$L_{\rm X}/L_{\rm bol}$}
\newcommand{\ra}{$\alpha$}
\newcommand{\dec}{$\delta$}
\newcommand{\pmra}{$\mu_{\alpha}$}
\newcommand{\pmdec}{$\mu_{\delta}$}
\newcommand{\rv}{$RV$}
\newcommand{\plx}{$\pi$}
\newcommand{\gimem}{{\it good initial members}}
\newcommand{\mo}{$M_{\rm 0}$}
\newcommand{\mnn}{$M_{\rm NN}$}

\newcommand{\mmp}{$M_{\rm MP}$}

\newcommand{\mfin}{$M_{\rm final}$}

\newcommand{\msun}{$M_{\odot}$}

\title[Development of models for NYMGs]{Development of models for nearby young stellar moving groups: creation, revision, and finalisation of the models}

\author[Lee \& Song]{
Jinhee Lee,$^{1}$\thanks{E-mail: jinhee@uga.edu}
and Inseok Song,$^{1}$
\\
$^{1}$The University of Georgia, Department of Physics and Astronomy, Athens, GA 30605, USA}

\date{Accepted XXX. Received YYY; in original form ZZZ}

\pubyear{2019}

\begin{document}
\label{firstpage}
\pagerange{\pageref{firstpage}--\pageref{lastpage}}
\maketitle

\begin{abstract}
Lee \& Song (2018, Paper 1) developed a tool for calculating Bayesian membership probability for nearby young stellar moving groups (BAMG: Bayesian Analysis of Moving Groups).
The study presented the importance of careful construction of models in moving group membership assessment, using $\beta$ Pictoris moving group as a test case. 
In this study, we build models for all major nearby young stellar moving groups (NYMGs hereafter) through 4 stages.
A set of prototype models is created using members listed in the discovery paper of each group.  
For each group, suggested members after the discovery of the group are used for revising these prototype models.
As these additional members being incorporated, group properties of a NYMG changes, thus membership probabilities change as well.
A subset of stars show consistently large membership probabilities regardless of the details of a chosen method for ingesting additional members.
Utilising these members, the NYMG models are finalised.  
The finalised models are applied for evaluating memberships of all claimed candidate members, resulting in a list of bona fide members. 
The mass function of bona fide members for the entire set of NYMG members indicates that more late-M type members should be identified.
 In addition, some objects showing a large difference in membership probability between BAMG and BANYAN $\Sigma$ \citet{gag18b} are presented and discussed.
Memberships of some planet host stars are changed, and it can have a significant influence on the estimated planetary masses.
\end{abstract}


%%%%%%%%%%%%%%%%%%%%%%%%%%%%%%%%%%%%%%%%%%%%%%%%%%

%%%%%%%%%%%%%%%%% BODY OF PAPER %%%%%%%%%%%%%%%%%%

%\begingroup
%\let\clearpage\relax
%\tableofcontents
%\endgroup

%\newpage

\section{Introduction}

%\begin{center}
%\input{table.tex}
%\end{center}

Nearby young stellar moving groups (NYMG hereafter) have attracted much attention since their discovery over the past two decades [e.g., \citet{kas97, web99, son03, zuc04b, tor08, rod11, zuc11, shk12, mal13, gag18b}], because NYMGs have potential applications in various stellar fields. 
NYMGs are loose stellar associations whose members formed together, and they are typically located within 100 pc and younger than 200 Myr. 
Due to both their youth and proximity to Earth they have been prime targets for infrared imaging of self-luminous exoplanets \citep{mar08}.  
They are also benchmarks for studying low-mass stars and brown dwarfs \citep{giz02} because their ages are relatively well constrained and their proximity  and youth make identification of their coolest members possible.  
This results in an improved mass function in the low mass regime \citep{kra14}. Additionally, NYMGs can allow us to understand star formation and evolution in a relatively low-density environment.  \newline

Because of their importance, the census of NYMGs has been a hot topic [e.g., \citet{zuc04b, tor08, shk11, mal13, ell16, shk17, gag18a}]. 
NYMG members must have formed together relatively recently in the same area, sharing similar motion.  
Therefore, probable members can be identified using signs of youth and their kinematic information in a 4 parameter space [i.e., R.A. (\ra) and Dec. (\dec) and proper motions (\pmra\ and \pmdec)]. 
However, Galactic position and Galactic velocity in the 6 parameter space ($XYZ$ and $UVW$) should provide the most comprehensive and accurate kinematic assessment.  
For this full position and velocity information, 6 parameters [i.e., \ra, \dec, \pmra, \pmdec, radial velocity (\rv), and parallax (\plx)] are required. 
In many cases, however, only a subset of those parameters are available (e.g., lacking \plx\ or \rv), especially for low mass stars. 
In such cases, instead of a unique value in the 6D space, a range of $XYZ$ and $UVW$ values can be calculated assuming an acceptable range of the missing parameters.
A Bayesian approach effectively deals with such cases through the marginalisation process \citep{gre05}. 
Due to this advantage and the fact that Bayesian posterior is based on current information, the Bayesian approach has grown in popularity for identifying NYMG members. \citet{mal13} and \citet{gag14a, gag18b} developed NYMG membership probability calculation tools based on the Bayesian framework. Under the presumed assumptions -- models of NYMGs and field stars --, these tools can provide a reliable membership probability for a given 
star, and the result is straightforward to interpret.  \newline

NYMGs were distinctive and easily separable from each other in XYZUVW when they were initially discovered (the late 1990s to the early 2000s). 
Later on, more and more NYMG members were identified, and new NYMGs were discovered as well [e.g., \citet{zuc01b, zuc04a, mam07, tor08, rod11, zuc11, shk12, mal13, gag18d}].  
The newly discovered members are identified from properties similar to previously identified members, and these additional members modify the group properties (i.e., mean position and velocity) in turn. 
Therefore, as many NYMG members are added, the distinctions among NYMGs have gotten less prominent causing overlapping distributions in XYZUVW, which raises confusions in NYMG memberships. 
Membership for a new candidate member is determined based on the average group property of the relevant NYMG. 
The group properties of a NYMG (e.g., mean positions and extensions in XYZ and UVW) are presented as model parameters, and these models can be constructed based on the previously identified members of the NYMG. Since false members can bias the model parameters, NYMG models should be carefully established using only the most reliable members. Thanks to the recently released Gaia DR2 \citep{gai18}, more NYMG members have \plx\ and \rv\ values, which allow more accurate membership assessment, enhancing the reliability of NYMG models.  \newline

In this study, we build up models of NYMGs through 4 stages beginning with members listed in the discovery paper of each NYMG. 
We describe the NYMG model establishment process in Section 2. 
In Section 3, the models and bona fide members are presented in a case of the $\beta$ Pictoris moving group. 
Models for the other NYMGs and their bona fide members are presented in Appendix. 
Discussion and conclusion appear in Section 4 and 5.

\section{Method}

In any NYMG membership calculation scheme, a set of model parameters such as average positions of members in XYZ and UVW for a NYMG need to be defined so that a candidate member's data can be examined against such parameters to assess its membership probability.
In a Bayesian scheme like ours and BANYAN variations, for a given NYMG, average position ($X_0$, $Y_0$, $Z_0$) and extension 
($\sigma_X$, $\sigma_Y$, $\sigma_Z$) of members in the 3D positional space (XYZ) are estimated fitting a 3D Gaussian ellipsoid.
To fully describe the best fit Gaussian ellipsoid in XYZ, in addition to six parameters ($X_0$, $Y_0$, $Z_0$, $\sigma_X$, $\sigma_Y$, $\sigma_Z$), we also need to specify its rotational orientation (i.e., Euler angles of the ellipsoid: $\phi_{\rm xyz}$, $\theta_{\rm xyz}$, $\psi_{\rm xyz}$).
Therefore, 9 parameters are necessary to describe an XYZ model for a given NYMG.
Similarly, another set of 9 parameters ($U_0$, $V_0$, $W_0$, $\sigma_U$, $\sigma_V$, $\sigma_W$, $\phi_{\rm uvw}$, $\theta_{\rm uvw}$, $\psi_{\rm uvw}$) is needed to describe NYMG's characteristics in the 3D velocity space UVW.
These 18 parameters can fully describe a NYMG in the 6D XYZUVW space, and we call them {\it NYMG model parameters}.
If a 6D ellipsoidal fitting to the distribution of members in XYZUVW is attempted, one needs to consider possible correlations among $XYZ$ and $UVW$.  However, in our approach, we treated $XYZ$ and $UVW$ as two independent sets of 3D parameters each of which can be fitted independently with a 3D ellipsoid.  
Therefore, we ignore cross-terms between $XYZ$ and $UVW$.
These NYMG model parameters are obtained utilising a Gaussian mixture model from the {\sc scikit-learn} package of {\sc python} \citep{ped11}. \newline

To build a model of a NYMG, a set of members is required; model parameters necessarily depend on the set of members.
Currently, claimed NYMG members overlap with those from other groups in XYZ and UVW, and this causes a significant overlap of models, 
hence an ambiguity in membership assessment.
On the other hand, at the time of discovery of a NYMG, the group was distinct from others.
More members have been claimed and model parameters were modified.
If false positive members are included in the input data set, the model would be distorted.
Once the model is distorted, the following identification of new members can be biased, which can lead to a misunderstanding 
of the properties of the NYMGs as well. \newline

\begin{figure*}
\centering
 \includegraphics[width=0.8\textwidth]{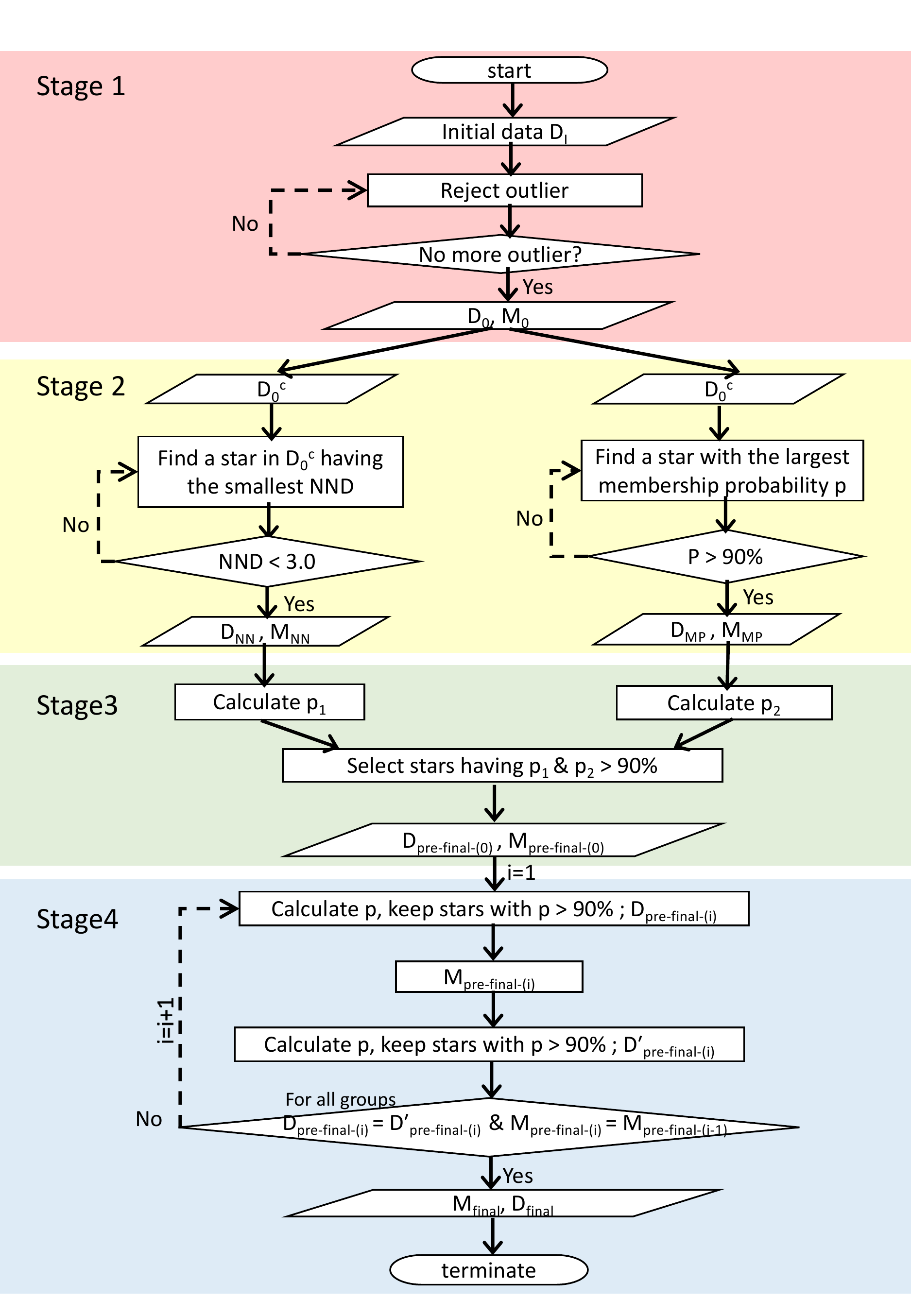}
 \caption{Flowchart of the process of developing NYMG models.
 The process consists of 4 stages: creation (Stage 1, in red), revision (Stage 2, in yellow), pre-finalisation (Stage 3, in green), and finalisation (Stage 4, in blue).
 $D$ represents a kinematic data set of stars, and subscripts describe classes of the data (e.g., $D_{\rm I}$, $D_0$, and $D_{\rm final}$ represent initial, good initial, and finalised members, respectively).
 $M$ represents a NYMG model (i.e., a set of 18 parameters), and subscripts indicate which set of stars are used to build the model (e.g., $M_0$ is generated using $D_0$).
 Following the convention of flow chart symbol, a diamond indicates a decision step, a parallelogram indicates input and output, and a rectangle indicates a  process.
The prototype models created in Stage 1 are revised in Stage 2 utilising two methods.
The two revised models (\mnn\ and \mmp) are used for selecting stars with large membership probabilities with these two models.
Using the selected stars, model finalisation process (Stage 3 and 4) is performed.
The details are explained in the text (Section 2.2-2.5).}
\label{fig:flowchart}
\end{figure*}

To prevent this potential issue, it is necessary to re-examine memberships of previously claimed members and establish NYMG models. 
The prototype models are created with the initial members that are listed in the discovery paper of the NYMG. 
However, the prototype models might be inadequate to calculate reliable membership probabilities, because the prototype model is based on only a subset of true members and hence does not represent the whole distribution of members. 
Therefore, the prototype models should be revised by incorporating appropriate supplementary members. 
In this study, the model establishment is performed through 4 stages. 
Initially, prototype models for all NYMGs are created utilising their initial members (Stage 1). 
One may accept that most of the initial members of a NYMG would be bona fide members of the group. 
However, some of the initial members might turn out to be false positives or outliers because updated data (e.g., measurement of \plx\ or \rv) can demonstrate that they are less associated with the group. 
Therefore, in Stage 1, these outliers are detected and eliminated, and a prototype model is created utilising the cleaned list of initial members. 
In Stage 2, the prototype model is revised by incorporating members identified after the discovery of the NYMG. 
From the set of these additional members, only appropriate members are taken instead of adopting all of the additional members. 
Then, in Stage 3 and 4, the revised model is finalised. 
Fig.~\ref{fig:flowchart} illustrates the process of building up NYMG models. 
Throughout this paper, $D$ indicates a kinematic data set of stars, and subscripts describe classes of the data (e.g., $D_I$, $D_0$, and $D_{\rm final}$ indicate initial, good initial, and finalised members, respectively). 
$M$ represents a NYMG model (i.e., a set of 18 parameters), and subscripts indicate which set of stars are used to build the model (e.g., $M_0$ is generated using $D_0$). 
Each stage is performed iteratively.
Details of each stage are described in Section 2.2$-$2.5.

\begin{landscape}

\begin{table}
\scriptsize
\setlength\tabcolsep{3pt} 
\begin{threeparttable}
\begin{tabular}{*{22}{c}}
\toprule
Name & SpT & \ra\ & \dec\  &  \pmra\ & \pmdec\ & \plx  & \rv & $B$ & $V$ & $G$ & $G\rm_{BP}$ & $G\rm_{RP}$ & $K$ & $W_1$ & $NUV$ & \lxlbol & EW(Li) & f(\mo)\tnote{a} & f(\mnn)\tnote{a}  & f(\mmp)\tnote{a} & f(\mfin)\tnote{a}  \\
& & hh:mm:ss & dd:mm:ss & mas yr$^{-1}$ & mas yr$^{-1}$ & mas & \kms & mag & mag & mag & mag & mag & mag & mag & mag & & \AA  \\
\hline
              2MASS J00011217+1535355 &        L4:  &  00:01:12.16 & +15:35:35.5 &  143.49 & -174.72 &   31.63 &       - &     - &     - &     - &     - &     - &     - &     - &     - &     - &    - &     0 &     0 &     0 &     0  \\ 
       2MASS J00040288-6410358        &           L1 &  00:04:02.88 & -64:10:35.8 &    64.0 &   -47.0 &       - &       - &     - &     - &     - &     - &     - & 14.01 & 13.41 &     - &     - &    - &     0 &     0 &     0 &     0  \\ 
       2MASS J00041589-8747254        &       M5.7 &  00:04:15.84 & -87:47:25.4 &   72.88 &   -38.1 &   19.84 &       - &     - &     - & 16.54 & 19.13 & 15.09 & 11.87 & 11.65 &     - &     - &    - &     0 &     0 &     0 &     0  \\ 
                        HD 105                         &      G0V   &  00:05:52.54 & -41:45:11.1 &   97.75 &  -76.42 &   25.74 &     2.2 &   8.1 &  7.51 &  7.38 &   7.7 &  6.94 &  6.12 &  6.07 &     - &  -4.4 & 140.0 &     0 &     4 &     4 &     4  \\ 
                         HD203                         &         F2   &  00:06:50.09 & -23:06:27.2 &   96.78 &  -47.12 &   25.02 &     9.7 &  6.55 &  6.17 &  6.08 &   6.3 &  5.77 &  5.24 &  5.13 &     - &  -5.3 & 87.0 &     0 &     2 &     2 &     2  \\ 
       2MASS J00065794-6436542        &        M9:  &  00:06:57.93 & -64:36:54.2 &   86.92 &  -61.66 &   23.26 &       - &     - &     - & 18.21 & 21.26 & 16.52 & 12.17 & 11.74 &     - &     - &    - &     0 &     0 &     0 &     0  \\ 
       2MASS J00111532-3756553        &       M5.7 &  00:11:15.33 & -37:56:55.3 &  100.75 &  -89.72 &   21.63 &       - &     - &     - & 15.91 & 18.45 & 14.47 & 11.21 & 11.02 &     - &     - &    - &     0 &     0 &     0 &     0  \\ 
       2MASS J00125703-7952073        &       M2.9 &  00:12:57.05 & -79:52:07.3 &   80.94 &  -45.12 &   21.42 &     9.4 & 15.06 & 13.56 & 12.41 & 13.81 & 11.26 &  8.75 &  8.66 & 19.86 &  -2.4 & 16.0 &     0 &     4 &     4 &     4  \\ 
                        HD 987                         &        G6V &  00:13:53.01 & -74:41:17.9 &   82.81 &  -49.08 &   21.81 &     9.2 &  9.48 &  8.73 &  8.62 &  9.04 &  8.09 &  6.96 &   6.9 & 14.55 &  -3.4 & 198.0 &     0 &     4 &     4 &     4  \\ 
                       HIP1134                        &         F5    &  00:14:10.25 & -07:11:56.8 &  104.54 &  -67.91 &   21.78 &     0.6 &  7.82 &  7.32 &  7.19 &  7.47 &   6.8 &  6.07 &  6.08 &     - &  -4.3 & 120.0 &     0 &   4,6 &   4,6 &     0  \\  
\multicolumn{22}{c}{\dots} \\
\multicolumn{22}{c}{\dots} \\
\multicolumn{22}{c}{\dots} \\
\bottomrule
\end{tabular}
\begin{tablenotes}
\item[a] These flags indicate whether the star is used in the construction of the corresponding model.
Numbers from 1 through 9 correspond to a NYMG: 1 (TWA), 2 (BPMG), 3 (ThOr), 4 (TucHor), 5 (Carina), 6 (Columba), 7 (Argus), 8 (ABDor), 9 (VCA). 
Zero indicates that the star is not used in building the model.
\end{tablenotes}
\end{threeparttable}
\caption{A sample of the claimed NYMG members. The entire table including membership probabilities [based on \mo, \mnn, \mmp, \mfin\ and from BANYAN $\Sigma$ (v1.2)] is available online.}
\label{tab:entiredata}
\end{table}
\end{landscape}

\begin{table*}
\begin{threeparttable}
\small
\setlength\tabcolsep{2pt} 
\begin{tabularx}{\textwidth}{C{0.15}L{0.15}C{0.06}L{0.15}C{0.04}L{0.4}}
\toprule
Name & Discovery  & Age & References for age & D\tnote{a} &  References for additional members \\
 (abbreviation)  & &   (Myr) & & (pc) \\
\hline 
The TW Hydrae association (TWA) & \citet{kas97}\tnote{b}  & 8-10 & \citet{tor08, bel15} & 50   &\citet{web99, ste99, zuc01a, giz02, rei03, son03, zuc04b, scz05, mam05, loo07, kir08, kas08, loo10a, loo10b, shk11, rod11, scn12b, mal13, rie14, gag14a, gag15a, gag15b, gag15c, mur15, ell16,  gag18b, gag18c}\\ \hline
The $\beta$ Pictoris moving group (BPMG)  & \citet{zuc01b}   &12-24 & \citet{zuc01b, bel15} & 35  &\citet{zuc04b, moo06, tor08, kir08, lep09, tei09, scl10, ric10, kis11, scl12a, scl12b, shk12, moo13, liu13, mal13, mal14a, mal14b, rie14, gag14a, rod14, gag15a, gag15b, bes15, bin15, ell16, shk17, gag18c} \\ \hline
The 32 Ori group (ThOr) & \citet{mam07}\tnote{c} & 20 & \citet{bel17} & 90  & \citet{bel17, gag18b, gag18c} \\ \hline
The Tucana/Horologium association (TucHor)  & \citet{zuc00, tor00}  & 30-45 & \citet{zuc04b, bel15} & 50& \citet{zuc00, zuc01d, zuc04b, moo06, tor08, kis11, zuc11, zuc12, shk12, rod13, mal13, del13, gag14a, mal14a, mal14b, kra14, gag15a, gag15b, art15, bin15, ell16, gag18b} \\ \hline
The Carina association (Carina)  & \citet{tor08}  & 30-45 & \citet{tor08, bel15} & 90 & \citet{shk12, mal13, moo13, mal14a, gag14a, gag15a, gag15b, bin15, ell16, sil16, gag18b, gag18c} \\ \hline
The Columba association (Columba) &  \citet{tor08}  & 30-42 & \citet{tor08, bel15} & 80   & \citet{zuc11, zuc12, shk12, mal13, moo13, rod13, gag14a, mal14a, mal14b, gag15a, gag15b, bin15, ell16, gag18b, gag18c} \\ \hline
The Argus association (Argus) & \citet{tor03a}   & 40 & \citet{tor08, bel15} & 97 & \citet{rie11, zuc11, zuc12, mal13, desil13, moo13, gag14a, mal14a, mal14b, rie14, gag14b, gag15a, gag15b, bes15}\\ \hline
The AB Doradus moving group (ABDor) & \citet{zuc04a}   & 50-150 & \citet{zuc04a, bel15} &  39 & \citet{lop06, tor08, scl10, zuc11, wah11, scl12a, scl12b, shk12, fah13, mal13, moo13, rod13, gag14a, mal14a, mal14b, rie14, gag15a, gag15b, bes15, bin15, ell16, des18, gag18b, gag18c} \\ \hline
Volans-Carina (VCA) &  \citet{oh17, gag18d}\tnote{d}   & 90 & \citet{gag18d} & 85  &  -- \\
\bottomrule
\end{tabularx}
\begin{tablenotes}
\item[a] \label{a} Mean distance of the group.
\item[b] \label{b} Since the number of discovery members is small (N=5), we included 6 systems successively identified by \citet{web99} in Stage 1.
\item[c] \label{c} Since a list of the discovery members is not presented in the paper \citep{mam07}, we took initial members from a successive work by \citet{bel17}.  "Known members" in \citet{bel17} are considered as initial members in Stage 1.
\item[d] \label{d} This group is discovered by \citet{oh17} as an ensemble of comoving stars.  \citet{fah18} identified this ensemble as a new coeval association with an age similar to the Pleiades, and a list of members is successively complied by \citet{gag18d}. 
We consider stars listed in \citet{gag18d} as initial members.
\end{tablenotes}
\caption{NYMGs considered in this study.}
\label{tab:groups}
\end{threeparttable}
\end{table*}

\clearpage

\subsection{Input data}

In this study, NYMGs having a mean distance of $\leq$100 pc and age of $\leq$200 Myr are considered (listed in Table~\ref{tab:groups}).
For detailed information about these groups, see Appendix A.

Table~\ref{tab:entiredata} lists approximately 2000 members of the NYMGs collected from references listed in Table~\ref{tab:groups}.
Members listed in the discovery paper of a NYMG are referred to as initial members, while members identified afterwards are referred to as additional members.
To present a star's position and motion in the 3D Galactic spaces (XYZ and UVW), 6 astrometric parameters (\ra, \dec, \pmra, \pmdec, \plx, and \rv) are required.
We used Gaia DR2 \citep{gai18} as our primary source for obtaining astrometric parameters.
For about 5 per cent of the collected NYMG members ($\sim$100 stars), \ra, \dec, \pmra, \pmdec\ are taken from other than Gaia DR2.
These stars are mostly late M or L type stars. 
The 2MASS All-Sky Point Source Catalog [2MASS; \citet{cut03}], the Tycho-2 catalogue [TYC2; \citet{hog00}], the fourth US naval observatory CCD astrograph catalogue [UCAC4; \citet{zac13}], and the Hipparcos catalogue [HIP; \citet{per97}] are used alternative sources of \ra, \dec\ or \pmra, \pmdec\ for these stars.
Regarding \plx, most data are collected from Gaia DR2, and data for 50 stars are available from the literature.
For about 100 stars (mostly late M or L type stars), no \plx\ is available to collect.
For \rv, Gaia DR2 has measurements for 360 stars.
For stars missing \rv\ in Gaia DR2, we considered RAVE 5 [The Radial Velocity Experiment: Fifth Data Release; \citet{kun17}], then \citet{tor06} and our internal database.
After searching literature, 800 stars are missing \rv.   \newline

We collected various age relevant data for all the claimed members from the literature and major catalogues, then evaluated their ages following the common age-dating methods (positions on colour-magnitude diagrams, NUV excess, \lxlbol, and Lithium 
$\lambda$6708$\AA$). When none of these methods is available (e.g., late M- or L-type members lacking \plx), we accept the 
youth evaluation of the original reference.

Stars with confirmed youth and full astrometric parameters are used as input data ($D$) in the model development process.

\begin{figure*}
\begin{subfigure}{0.3\textwidth}
\includegraphics[width=\textwidth]{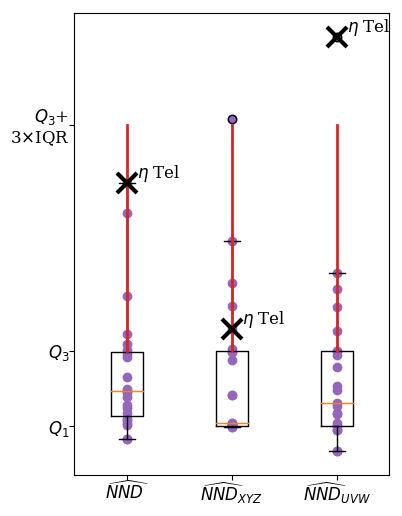} 
\end{subfigure}
\begin{subfigure}{0.69\textwidth}
\includegraphics[width=\textwidth]{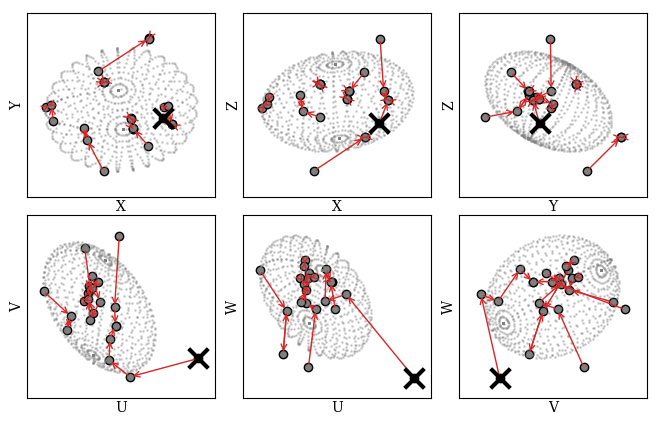}
\end{subfigure}
\caption{The first iteration of the outlier elimination process utilising the nearest neighbour distance ($NND$).
An outlier in this iteration is marked with a cross (the name is denoted in the left panel).
Left: boxplots of distributions of re-scaled $NND$ in XYZ, UVW, and the mean values ($\widehat{NND}$, $\widehat{NND}_{\rm xyz}$, and $\widehat{NND}_{\rm uvw}$, respectively).
Boxes denote the range from first quartile ($Q_1$) to third quartile ($Q_3$), and red lines stretch from $Q_3$ to $Q_3 + 3\times IQR$ (interquartile range).
$\eta$ Tel is rejected due to the position in $UVW$ (beyond the tip of the red line).
Right: each star is connected to its nearest neighbour with an arrow.
The temporary ellipsoid model (2$\sigma$) fitted with the presented data (excluding the outlier) is presented with grey colour.}
\label{fig:nn_process}
\end{figure*}

\subsection{Stage 1: creation of prototype models}

This process begins with a set of initial members of a NYMG ($D_I$), which is a subset of the full input data $D$.
As illustrated in Fig.~\ref{fig:flowchart} and described at the beginning of Section 2, outliers are eliminated via iteration. \newline

In this set of initial members, we start by eliminating outliers (e.g., false members). 
Bayesian membership probability \citep{mal13, gag14a, lee18} is a straightforward indicator of memberships.
However, our Bayesian membership probability calculation requires models of all considered NYMGs, which have not been created yet at the beginning.

An alternative outlier rejection method is to use nearest neighbour distance ($NND$ hereafter).
A star having the largest $NND$ (i.e., separated from other stars) can be assessed as an outlier.
For a star, $NND$ is calculated in XYZ and UVW ($NND_{\rm xyz}$ and $NND_{\rm uvw}$).
To identify which star is the most isolated one (i.e., outlier), the $NND_{\rm xyz}$ and $NND_{\rm uvw}$ are averaged so that each star has one single $NND$ value.
Because $XYZ$ and $UVW$ have different scales, $NND_{\rm xyz}$ and $NND_{\rm uvw}$ should be normalised first.
We use interquartile range ($IQR_{\rm xyz}$) and 3rd quartile (Q$_{\rm 3,xyz}$) values to rescale $NND_{\rm xyz}$.

\begin{equation}
\widehat{NND}_{\rm xyz} = \frac{NND_{\rm xyz}-Q_{\rm 3,xyz}}{IQR_{\rm xyz}}
\end{equation}

In UVW, $NND_{\rm uvw}$ can be normalised as the same way as $NND_{\rm xyz}$.
The averaged $NND$ value ($\widehat{NND}$) for a star can be obtained by the arithmetic mean of $\widehat{NND}_{\rm xyz}$ and $\widehat{NND}_{\rm uvw}$.

We accept the outlier definition by \citet{tuk79} [samples with data of $>$ $Q_3$ + 3$\times IQR$ or $<$ $Q_1$ - 3$\times IQR$ are defined as outliers ({\it far out})].
In Stage 1, outliers from the initial members ($D_I$) are eliminated via iteration until no outlier is detected.
The mean $NND$ value ($\widehat{NND}$) of stars would be used for evaluating outliers (i.e., stars having $\widehat{NND}\geq$3.0 are defined as outliers).
However, in Stage 1, we use both $\widehat{NND}_{\rm xyz}$ and $\widehat{NND}_{\rm uvw}$ values in evaluation of outliers because the number of the initial members is small and one eccentric member in XYZ and UVW can significantly distort the distribution of members.
In Stage 1, $\widehat{NND}$ is used for sorting members to evaluate the most isolated member.
Then, if the star has the rescaled $NND$ value larger than 3.0 in XYZ or UVW, the star is eliminated.

\begin{figure*}
\includegraphics[width=0.99\textwidth]{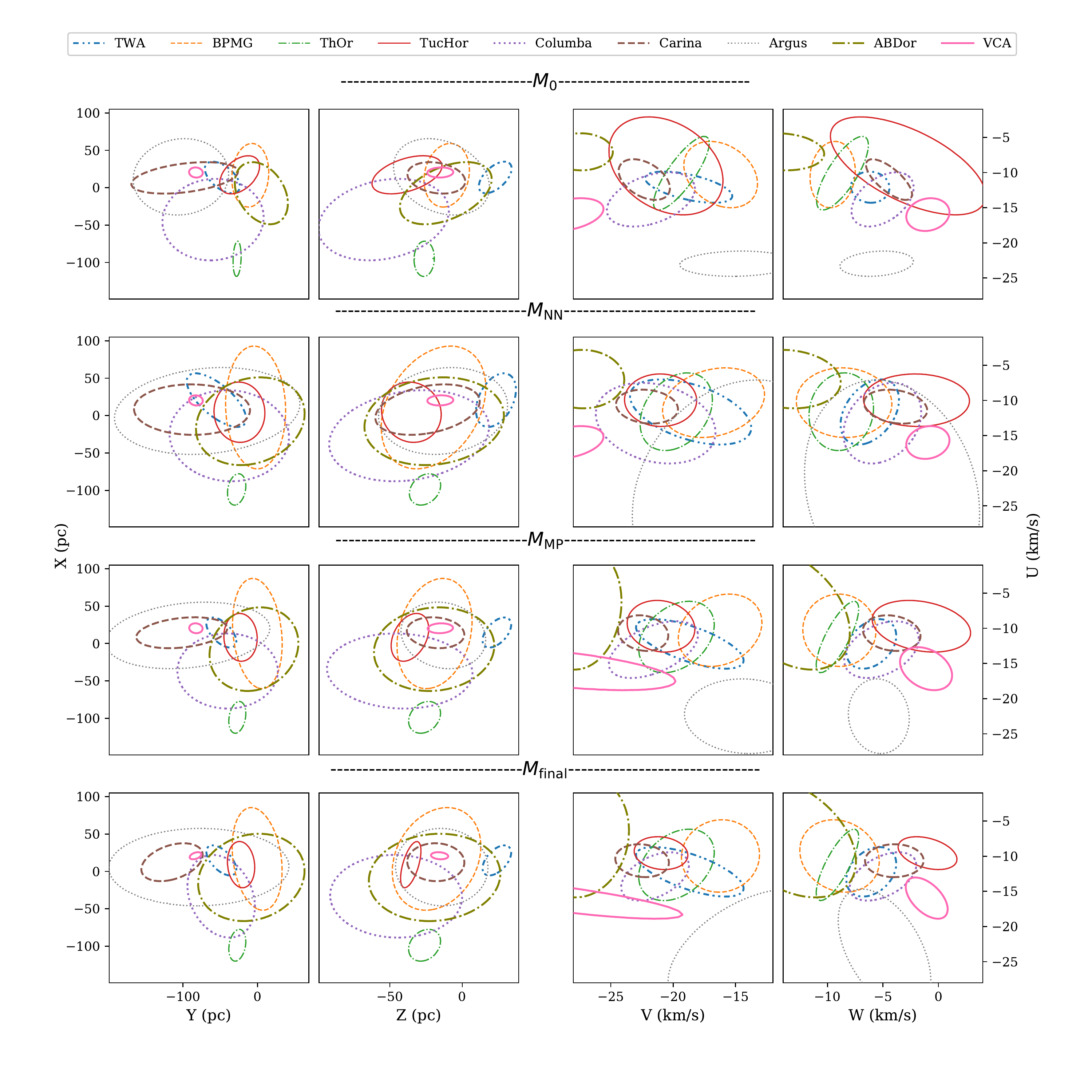}
\caption{The prototype, revised, and finalised NYMG models are illustrated in 2D spaces.
Models from top to bottom: prototype model (\mo), revised model using Method 1 (\mnn), revised model using Method 2 (\mmp), and finalised model (\mfin).}
\label{fig:mdls}
\end{figure*}

Fig.~\ref{fig:nn_process} illustrates the first iteration of detecting an outlier for BPMG.
In the figure, the outlier is marked with a large cross.
The $\widehat{NND}$, $\widehat{NND}_{\rm xyz}$, and $\widehat{NND}_{\rm uvw}$ values are displayed in the left panel of the figure.
In the figure, the tip of the red line indicates the boundary of outlier determination (i.e., 3.0).
In this first turn, $\eta$ Tel is evaluated as the most isolated member because the star has the largest $\widehat{NND}$ value as can be seen in the figure.
Even though $\widehat{NND}$ of $\eta$ Tel is not larger than 3.0, the star is evaluated as an outlier due to $\widehat{NND}_{\rm uvw}$ being larger than 3.0.
The remaining members in this turn are displayed as large grey dots in the right panel of Fig.~\ref{fig:nn_process}.
The outlier ($\eta$ Tel) in this turn is marked with a big cross.
Each star is connected to its nearest neighbour with a red arrow.
A temporary ellipsoidal model excluding the outlier is illustrated with small grey dots.
Next iteration begins with the data set excluding this outlier.
The remaining members after the termination of Stage 1 are referred to as \gimem\ ($D_0$).
They are fitted with a 3D Gaussian mixture model in XYZ and UVW to obtain the prototype model (\mo).
The model parameters of \mo\ are shown in Table~\ref{tab:mdls}, and prototype models \mo\ are illustrated in the top panel of Fig.~\ref{fig:mdls}.

\begin{table*}
\centering
\small
%\scriptsize
\setlength\tabcolsep{3pt} 
\begin{threeparttable}
\begin{tabular}{crrrrrrrrrrrrrrrrrrr}
\toprule
Name & $X$ & $Y$ & $Z$ & $\sigma_{\rm X}$ & $\sigma_{\rm Y}$ & $\sigma_{\rm Z}$ & $\phi_{\rm xyz}$ & $\theta_{\rm xyz}$ & $\psi_{\rm xyz}$ & $U$ & $V$ & $W$ & $\sigma_{\rm U}$ & $\sigma_{\rm V}$ & $\sigma_{\rm W}$ & $\phi_{\rm uvw}$ & $\theta_{\rm uvw}$ & $\psi_{\rm uvw}$ & $N$ \\
\cmidrule(lr){2-4}\cmidrule(lr){5-7}\cmidrule(lr){8-10}\cmidrule(lr){11-13}\cmidrule(lr){14-16}\cmidrule(lr){17-19} 
& \multicolumn{3}{c}{pc} & \multicolumn{3}{c}{pc} & \multicolumn{3}{c}{$\circ$} & \multicolumn{3}{c}{\kms} & \multicolumn{3}{c}{\kms} & \multicolumn{3}{c}{$\circ$} \\
\hline 
\multicolumn{20}{c}{\mo} \\ \hline
      TWA &  13.7 & -47.2 &  22.8 &  14.7 &   7.5 &   3.3 & -128 &  -18 &   -7 & -12.1 & -18.8 &  -6.1 &   1.9 &   0.9 &   0.7 & -115 &   -5 &  -46 &  15  \\ 
      BPMG &  16.7 &  -9.7 & -10.7 &  21.4 &  12.7 &   7.1 &  175 &   -3 &  -17 & -10.3 & -16.2 &  -9.5 &   2.4 &   1.4 &   1.0 & -165 &   -2 &   12 &  21  \\ 
      ThOr & -95.1 & -27.4 & -26.3 &  11.8 &   4.1 &   1.6 &  177 &    0 &   55 & -10.1 & -19.3 &  -8.6 &   2.9 &   0.9 &   0.3 &  160 &  -18 &  126 &  13  \\ 
    TucHor &  17.3 & -24.0 & -38.2 &  19.1 &  10.5 &   3.3 &  124 &  -36 &   20 &  -9.1 & -20.6 &  -2.7 &   4.4 &   2.6 &   1.7 & -172 &   44 &  -47 &  30  \\ 
    Carina &  13.1 & -97.7 & -17.9 &  36.2 &  10.1 &   9.0 &   97 &    4 &   43 & -11.0 & -22.3 &  -4.5 &   1.8 &   0.9 &   0.6 & -155 &   30 &   -3 &  21  \\ 
   Columba & -42.6 & -59.7 & -54.3 &  34.1 &  28.2 &  21.1 &   98 &   -5 &  -23 & -13.8 & -21.8 &  -5.0 &   2.3 &   1.4 &   1.2 &  141 &  -22 &   -5 &  35  \\ 
     Argus &  14.5 & -102.8 & -14.4 &  32.2 &  26.1 &  15.4 &   97 &   -3 &   16 & -23.0 & -14.9 &  -5.6 &   2.5 &   1.3 &   0.9 &   93 &  -28 &  -77 &  20  \\ 
     ABDor &  -7.3 &   5.1 & -11.4 &  23.7 &  17.2 &  11.9 & -153 &  -23 & -144 &  -7.1 & -27.4 & -14.1 &   2.0 &   1.3 &   1.2 &  113 &   72 &   -4 &  34  \\ 
       VCA &  20.4 & -82.6 & -15.2 &   4.8 &   4.3 &   3.3 &  -91 &   37 &   78 & -16.0 & -28.4 &  -0.9 &   1.5 &   1.1 &   0.9 &  121 &    1 &  -40 &  18  \\ 
     \hline
\multicolumn{20}{c}{\mnn} \\ \hline
      TWA &  20.7 & -55.1 &  23.3 &  24.9 &  10.4 &   6.3 & -130 &   -6 &    4 & -11.7 & -18.6 &  -6.2 &   3.0 &   1.7 &   1.1 & -131 &  -14 &   -4 &  40  \\ 
      BPMG &  10.9 &  -2.6 & -19.9 &  41.6 &  20.5 &  16.4 & -177 &   -9 &  161 & -10.3 & -16.8 &  -8.5 &   2.7 &   2.3 &   1.5 &  141 &   26 &  -46 & 145  \\ 
      ThOr & -98.7 & -28.0 & -25.7 &  10.8 &   7.0 &   3.4 &  166 &  -12 &   36 & -11.6 & -19.8 &  -8.8 &   2.8 &   1.5 &   1.3 &  167 &   -4 &   55 &  34  \\ 
    TucHor &   4.5 & -24.3 & -35.2 &  20.2 &  17.6 &   9.3 & -172 &    4 &   16 & -10.0 & -21.0 &  -2.0 &   2.4 &   1.9 &   1.4 &   72 &  -80 &  -17 & 226  \\ 
    Carina &   8.0 & -87.9 & -24.0 &  39.1 &  19.9 &  14.1 &  -91 &   -6 & -131 & -10.9 & -22.1 &  -3.9 &   1.7 &   1.1 &   1.0 & -120 &   50 &   15 &  67  \\ 
   Columba & -26.9 & -37.9 & -36.7 &  42.2 &  31.5 &  22.9 &   91 &  -22 &  158 & -13.2 & -21.4 &  -5.1 &   3.0 &   2.3 &   1.6 & -156 &  -10 &   20 & 143  \\ 
     Argus &   6.2 & -67.5 & -13.2 &  62.7 &  28.7 &  21.3 &   95 &   -5 &   -7 & -23.2 & -14.9 &  -4.2 &   8.1 &   4.5 &   3.4 &  171 &    7 &  -39 &  85  \\ 
     ABDor &  -7.5 &  -9.7 & -19.4 &  38.9 &  28.2 &  21.5 &  108 &  -19 &    3 &  -7.0 & -27.3 & -13.6 &   2.4 &   2.0 &   1.7 & -173 &   70 &   84 & 152  \\ 
       VCA &  20.4 & -82.6 & -15.2 &   4.8 &   4.3 &   3.3 &  -91 &   37 &   78 & -16.0 & -28.4 &  -0.9 &   1.5 &   1.1 &   0.9 &  121 &    1 &  -40 &  18  \\ 
\hline  
\multicolumn{20}{c}{\mmp} \\ \hline
      TWA &  14.7 & -48.5 &  23.9 &  13.1 &   6.5 &   3.6 & -134 &  -15 &   -6 & -12.2 & -18.7 &  -6.0 &   2.6 &   1.2 &   0.9 & -125 &  -14 &  -27 &  31  \\ 
      BPMG &  13.3 &  -0.2 & -19.1 &  37.2 &  18.2 &   9.6 & -173 &   -5 &  -31 & -10.3 & -16.2 &  -8.9 &   2.6 &   1.8 &   1.5 &  166 &    2 &  -48 & 120  \\ 
      ThOr & -98.8 & -27.1 & -26.0 &  11.0 &   7.1 &   2.3 &  166 &  -12 &   42 & -11.2 & -19.7 &  -9.1 &   2.7 &   1.4 &   0.6 &  164 &  -16 &   -1 &  31  \\ 
    TucHor &   8.2 & -22.8 & -36.1 &  16.2 &  11.4 &   5.5 & -177 &   -9 &   13 &  -9.7 & -21.0 &  -1.5 &   2.4 &   1.7 &   1.3 & -151 &   60 &   59 & 186  \\ 
    Carina &  14.4 & -101.6 & -18.4 &  30.7 &  10.1 &   9.5 &   96 &    4 &   37 & -10.7 & -22.4 &  -4.2 &   1.4 &   1.2 &   0.9 & -149 &   47 &   58 &  46  \\ 
   Columba & -36.8 & -40.1 & -43.2 &  36.4 &  25.2 &  20.9 &  -91 &   27 &   -9 & -13.0 & -21.6 &  -5.1 &   2.5 &   1.5 &   1.3 &  143 &  -23 &  113 & 106  \\ 
     Argus &  10.8 & -93.8 & -13.1 &  55.7 &  22.1 &  13.0 &   95 &   -5 &   13 & -22.5 & -14.2 &  -5.4 &   2.7 &   2.4 &   1.3 & -159 &    5 &    8 &  52  \\ 
     ABDor &  -7.6 &  -4.8 & -19.4 &  31.9 &  26.6 &  19.4 &  120 &  -15 &    8 &  -7.2 & -27.5 & -13.7 &   4.6 &   2.5 &   1.6 &  176 &   21 &  -78 & 142  \\ 
       VCA &  20.4 & -82.7 & -15.1 &   4.7 &   4.2 &   3.3 &  -91 &   36 &   78 & -15.7 & -29.5 &  -1.1 &   4.9 &   1.2 &   1.0 & -100 &    7 &    3 &  19  \\ 
\hline  
\multicolumn{20}{c}{\mfin} \\ \hline
       TWA &  14.7 & -48.5 &  23.9 &  13.1 &   6.5 &   3.6 & -134 &  -15 &   -6 & -12.2 & -18.7 &  -6.0 &   2.6 &   1.2 &   0.9 & -125 &  -14 &  -27 &  30  \\ 
      BPMG &  16.7 &  -0.5 & -17.8 &  34.9 &  18.5 &  11.5 & -170 &   -8 &  -38 & -10.0 & -16.2 &  -8.9 &   2.6 &   1.9 &   1.4 &  176 &   11 &  -57 & 113  \\ 
      ThOr & -98.8 & -27.1 & -26.0 &  11.0 &   7.1 &   2.3 &  166 &  -12 &   42 & -11.2 & -19.7 &  -9.1 &   2.7 &   1.4 &   0.6 &  164 &  -16 &   -1 &  31  \\ 
    TucHor &   9.1 & -22.1 & -35.6 &  15.7 &   9.1 &   2.8 & -173 &   -8 &   -1 &  -9.6 & -21.0 &  -1.0 &   1.7 &   1.1 &   0.6 & -125 &   49 &   13 & 160  \\ 
    Carina &  12.3 & -114.7 & -18.3 &  21.6 &  11.5 &   9.2 &  108 &    8 &  -18 & -10.6 & -22.5 &  -4.0 &   1.5 &   1.2 &   0.9 &  -99 &   57 &   20 &  37  \\ 
   Columba & -33.3 & -48.8 & -45.7 &  29.2 &  24.1 &  19.1 & -151 &   12 &   53 & -12.9 & -21.5 &  -5.0 &   2.1 &   1.2 &   1.1 &  149 &  -31 &  106 &  87  \\ 
     Argus &   5.7 & -78.7 & -14.2 &  60.5 &  25.9 &  15.9 &  -89 &    1 &   -1 & -23.4 & -14.0 &  -4.9 &   4.9 &   2.6 &   1.7 &  151 &   12 &   22 &  46  \\ 
     ABDor &  -8.1 &  -8.3 & -19.1 &  37.6 &  28.4 &  20.7 &  107 &  -17 &    4 &  -7.3 & -27.4 & -13.6 &   4.6 &   2.6 &   1.9 &  176 &   25 &  -91 & 129  \\ 
       VCA &  20.9 & -83.2 & -15.7 &   4.1 &   3.0 &   1.9 &  -71 &    8 &  -72 & -16.0 & -29.6 &  -1.1 &   5.4 &   0.9 &   0.6 & -103 &    7 &   15 &  19  \\       
\bottomrule
\end{tabular}
\end{threeparttable}
\caption{The prototype (\mo), revised (\mnn\ and \mmp), and finalised (\mfin) model parameters for NYMGs.
$X, Y, Z, U, V$, and $W$ are centre positions of the models.
$\sigma$ values are the standard deviation of the models, and angles represent Euler angles.
The members used in building the models can be found in Table~\ref{tab:entiredata}, and $N$ indicates the number of these members.}
\label{tab:mdls}
\end{table*}

\subsection{Stage 2: revision of the prototype models}

As explained in Section 2, prototype models are based on a small group of initially discovered members hence they need to be updated by incorporating additionally suggested members.
In the set of input data ($D$), the set of members used to build the prototype models in Stage 1 is $D_0$, and the complement of the set $D_0$ are referred to as $D_0^c$.
This set $D_0^c$ consists of stars identified after the discovery of the group (additional members).
When model parameters are changed, some marginally rejected outliers from Stage 1 may become acceptable members.
Therefore, we included all rejected stars from $D_I$ in $D_0^c$ as well.
In $D_0^c$, appropriate members are selected and incorporated into $D_0$ to revise the set of members (referred to as $D_{\rm rev}$).
This revision processes (i.e., member addition) is performed iteratively.
The iteration is terminated when there are no more members to consider in $D_0^c$.
After the termination of the process, the revised model ($M_{\rm rev}$) is constructed using $D_{\rm rev}$. \newline

When evaluating stars in $D_0^c$, in addition to the $NND$ approach as in the creation of prototype models, Bayesian membership probability can be used as well because models of NYMGs were created already in Stage 1 (the prototype models \mo).
Therefore, we also utilise membership probability to evaluate appropriate members as an independent method in Stage 2.
These two methods [utilising nearest neighbour distance (NN, hereafter) and membership probability (MP, hereafter)] are used independently in Stage 2.
This process is briefly illustrated in Fig.~\ref{fig:flowchart}.

\subsubsection{Method 1: utilising nearest neighbour distance (NN)}

At the beginning, the set of members for revision ($D_{\rm rev}$) is identical to $D_0$, while the complement of the set $D_{\rm rev}$ (i.e., $D_{\rm rev}^c$) is identical to $D_0^c$ ($D_{\rm rev} = D_0$ and $D_{\rm rev}^c = D_0^c$).
In each iteration, an appropriate member $s_i$ in $D_{\rm rev}^c$ is evaluated and determined if it should be added to $D_{\rm rev}$.
The following describes the sequence of an iteration for evaluating $s_i$.

\begin{enumerate}
\item For $s_1$, the nearest neighbour in $D_{\rm rev}$ is found and distances to the neighbour ($NND_{\rm xyz,1}$ and $NND_{\rm uvw,1}$) are calculated.
This is performed for all stars in $D_{\rm rev}^c$ ($s_2, s_3, ..., s_n$).
\item These $NND$ values are re-scaled in each XYZ and UVW, and averaged ($\widehat{NND}_1$, $\widehat{NND}_2$, ... $\widehat{NND}_n$). 
\item The star having the smallest $\widehat{NND}$ is evaluated as the best appropriate member in this iteration.
\item If the star is not an outlier ($\widehat{NND}<$3), the star is moved from $D_{\rm rev}^c$ to $D_{\rm rev}$.
Otherwise, the iteration is terminated.
\end{enumerate}

Once the iteration is terminated, the data set $D_{\rm rev}$ is referred to as $D_{\rm NN}$, and the model revision is performed using stars in $D_{\rm NN}$. 
Table~\ref{tab:mdls} and the second row in Fig.~\ref{fig:mdls} present these revised models (\mnn).

\subsubsection{Method 2: applying the Bayesian membership probability (MP)}

Same as Method 1 in Section 2.3.1, at the beginning, the set of members for revision ($D_{\rm rev}$) is identical to $D_0$, while the complement of the set $D_{\rm rev}$ is identical to $D_0^c$ ($D_{\rm rev} = D_0$ and $D_{\rm rev}^c = D_0^c$).
In each iteration, an appropriate member $s_i$ in $D_{\rm rev}^c$ is evaluated and determined if it should be added to $D_{\rm rev}$.
The following describes the sequence of an iteration for evaluating $s_i$.

\begin{enumerate}
\item NYMG models are constructed ($M_{\rm rev}$) utilising members in $D_{\rm rev}$.
\item Bayesian membership probabilities for stars in $D_{\rm rev}$ and for those in $D_{\rm rev}^c$ are calculated.
\item In $D_{\rm rev}^c$, the star $s_i$ having the largest membership probability is evaluated as the best appropriate member in this iteration.
\item If the star $s_i$ has membership probability below the threshold ($p_{\rm cut}$=90 per cent), the iteration is terminated.
Otherwise, the star $s_i$ is moved from $D_{\rm rev}^c$ to $D_{\rm rev}$.
If there are members having membership probability smaller than $p_{\rm cut}$ in $D_{\rm rev}$, they are eliminated from $D_{\rm rev}$.
 \end{enumerate}

Once the iteration is terminated, the data set $D_{\rm rev}$ is referred to as $D_{\rm MP}$, and the model revision is performed (\mmp) using stars in $D_{\rm MP}$. \newline

Our determination of $p_{\rm cut}$ relies on the information in Stage 1.
More than 90 per cent of members in $D_0$ have Bayesian membership probabilities larger than 90 per cent excepting for TucHor.
For TucHor, the recovery rate is 0.8 with $p_{\rm cut}$ = 90 per cent, which means that about 80 per cent of TucHor stars have Bayesian membership probabilities larger than 90 per cent.
Therefore, we set the threshold to 90 per cent ($p_{\rm cut} \equiv$ 90 per cent).
The revised models (\mmp) are presented in Table~\ref{tab:mdls} and the third row in Fig.~\ref{fig:mdls}. \newline

\subsection{Stage 3: pre-finalisation of the models}
This stage is for preparation of the next stage. 
The two model revision methods in Stage 2 result in two different models for a given NYMG.
These two methods (hence their MG models) complement each other, and stars showing high membership probabilities regardless of which model is used  can be regarded as acceptable members.
In this stage, stars with $p>$90 per cent in both calculations based on \mnn\ and \mmp\ are selected.
For each NYMG, a set of these stars ($D_{\rm pre-final-(0)}$) is created.
Then, a model $M_{\rm pre-final-(0)}$ is constructed using $D_{\rm pre-final-(0)}$.

\subsection{Stage 4: finalisation of the models}

With $M_{\rm pre-final-(0)}$, now we can calculate membership probabilities of all candidate NYMG members and make a selection of acceptable members again with a certain threshold in membership probability (e.g., $>$90 per cent).
Then the final selected members can be used to modify the NYMG model once again.
This newly calculated NYMG model ($M_{\rm pre-final-(i+1)}$) can be different from the model in the previous stage ($M_{\rm pre-final-(i)}$) because some members could be rejected or new members were included through this recalculation.
Therefore, to make a self-consistent set of membership list and NYMG model, one needs to iterate through the process of (1) recalculation of membership based on $M_{\rm pre-final-(i)}$, (2) modify $D_{\rm pre-final-(i+1)}$, and (3) calculate $M_{\rm pre-final-(i+1)}$ until the iteration does not make any change in the membership list.
At the end of the iteration, we can create final membership list ($D_{\rm final}$) and model parameters ($M_{\rm final}$).

The finalised models are presented in Table~\ref{tab:mdls} and the bottom panel in Fig.~\ref{fig:mdls}.
Membership probability for the entire data set is calculated.
For a set of stars having membership probability of larger than 90 per cent with confirmed youth and full 6 astrometric parameters are referred to as {\it bona fide members}.
Stars having full 6 astrometric parameters with confirmed youth, but with slightly low membership probabilities (80$-$90 per cent) are referred to as {\it highly likely members}.
In the same set, stars with confirmed youth but lacking \rv\ or \plx\ are referred to as {\it probable members}, while stars with full 6 astrometric parameters but without age information are referred to as {\it possible members}.

\section{Results: a case of the $\beta$ Pictoris Moving Group}

In this section, using BPMG as an example, we compare \mnn, \mmp, \mfin\ and present the list of bona fide members of BPMG.
The results for other NYMGs are presented in Appendix B.

\begin{figure*}
\includegraphics[width=0.9\textwidth]{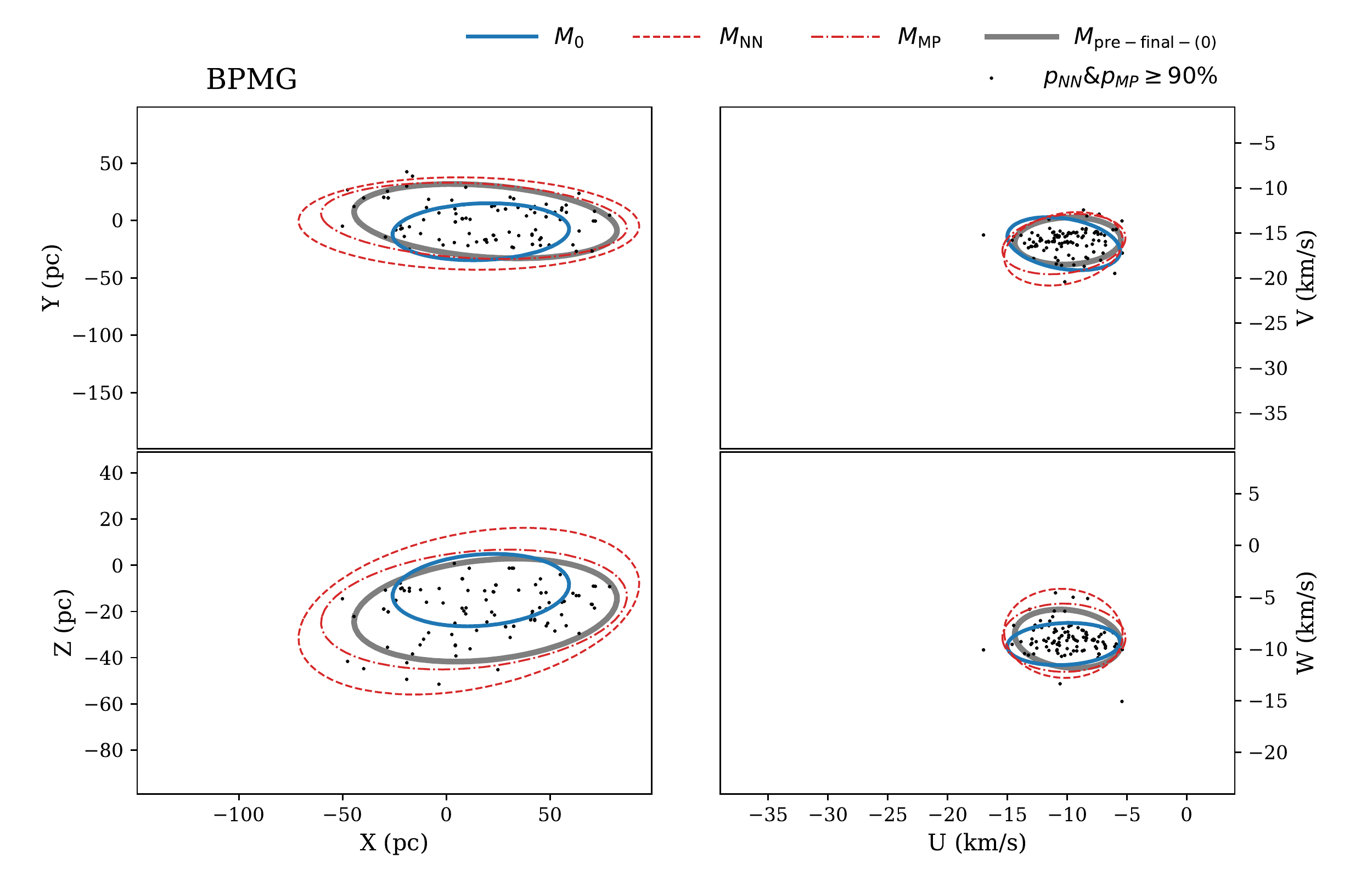}
\caption{The prototype, revised, and pre-finalised models for BPMG (2$\sigma$).
Blue solid, red dashed, red dot-dashed, and thick gray solid lines represent \mo, \mnn, \mmp, and $M\rm_{pre-final-(0)}$, respectively.
Models for other NYMGs are illustrated with the same range of axes (Figs.~\ref{fig:twa}$-$~\ref{fig:vca}).
Members with membership probabilities larger than 90 per cent based on \mnn\ and \mmp\ are displayed.
The $M\rm_{pre-final-(0)}$ model is created using these members.}
\label{fig:bpmg}
\end{figure*}

\begin{figure*}
\includegraphics[width=0.9\textwidth]{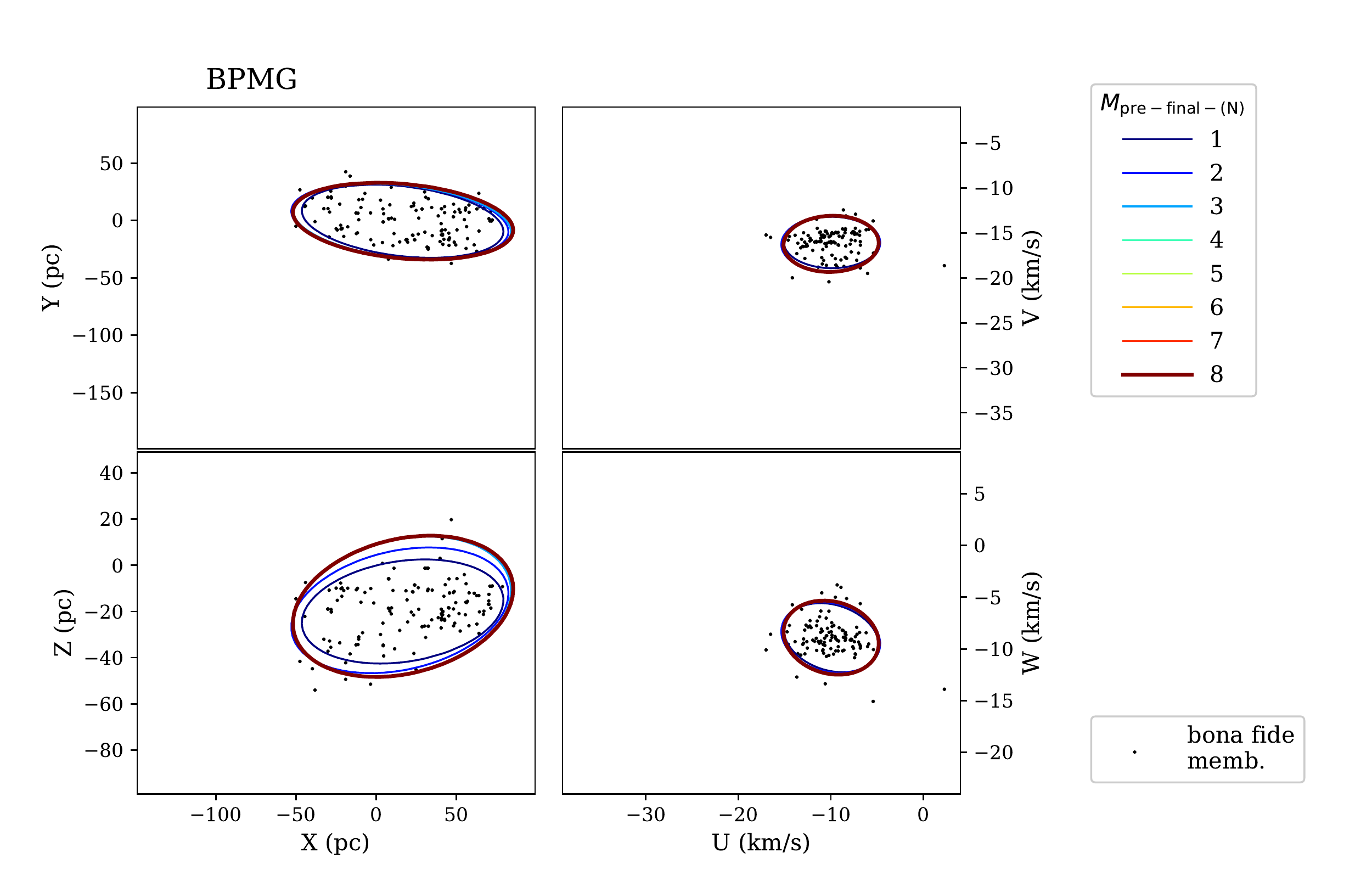}
\caption{Model finalisation process in Stage 4.
Models from the beginning ($M\rm_{pre-final-(1)}$) and to the last steps ($M\rm_{pre-final-(8)}$) are presented.
A model at the last step is the finalised model (\mfin) and is presented with a thick line.
Bona fide members are presented with small dots.}
\label{fig:bpmg2}
\end{figure*}

\begin{figure*}
\includegraphics[width=0.9\textwidth]{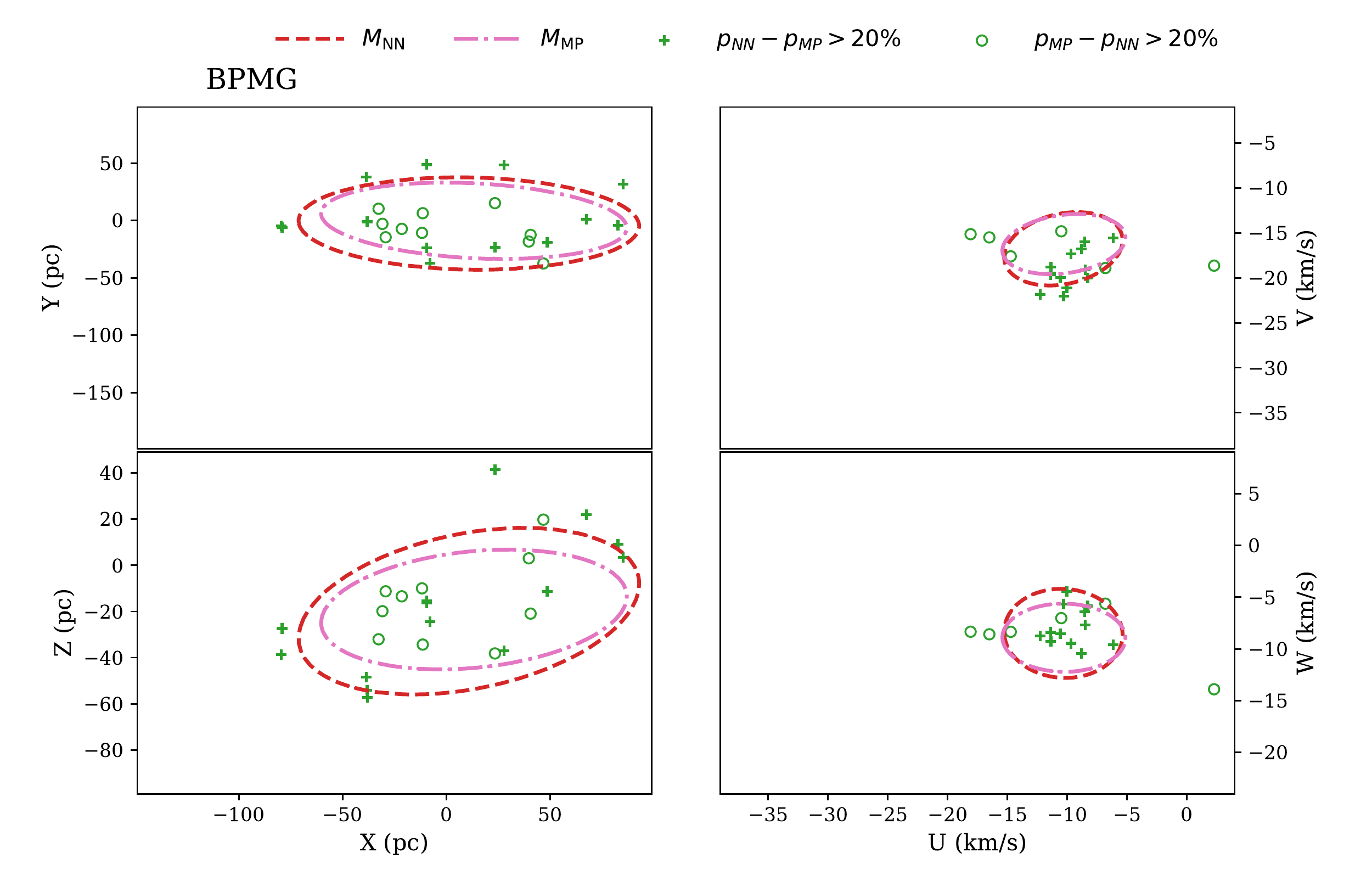}
\caption{Two different \mnn\ and \mmp\ models can cause a significant membership probability difference for some members.
These members ($\Delta p>$20 per cent) are displayed.
These stars are presented in Fig.~\ref{fig:bpmg_prob} as well.}
\label{fig:bpmg3}
\end{figure*}

\begin{figure*}
\includegraphics[width=0.9\textwidth]{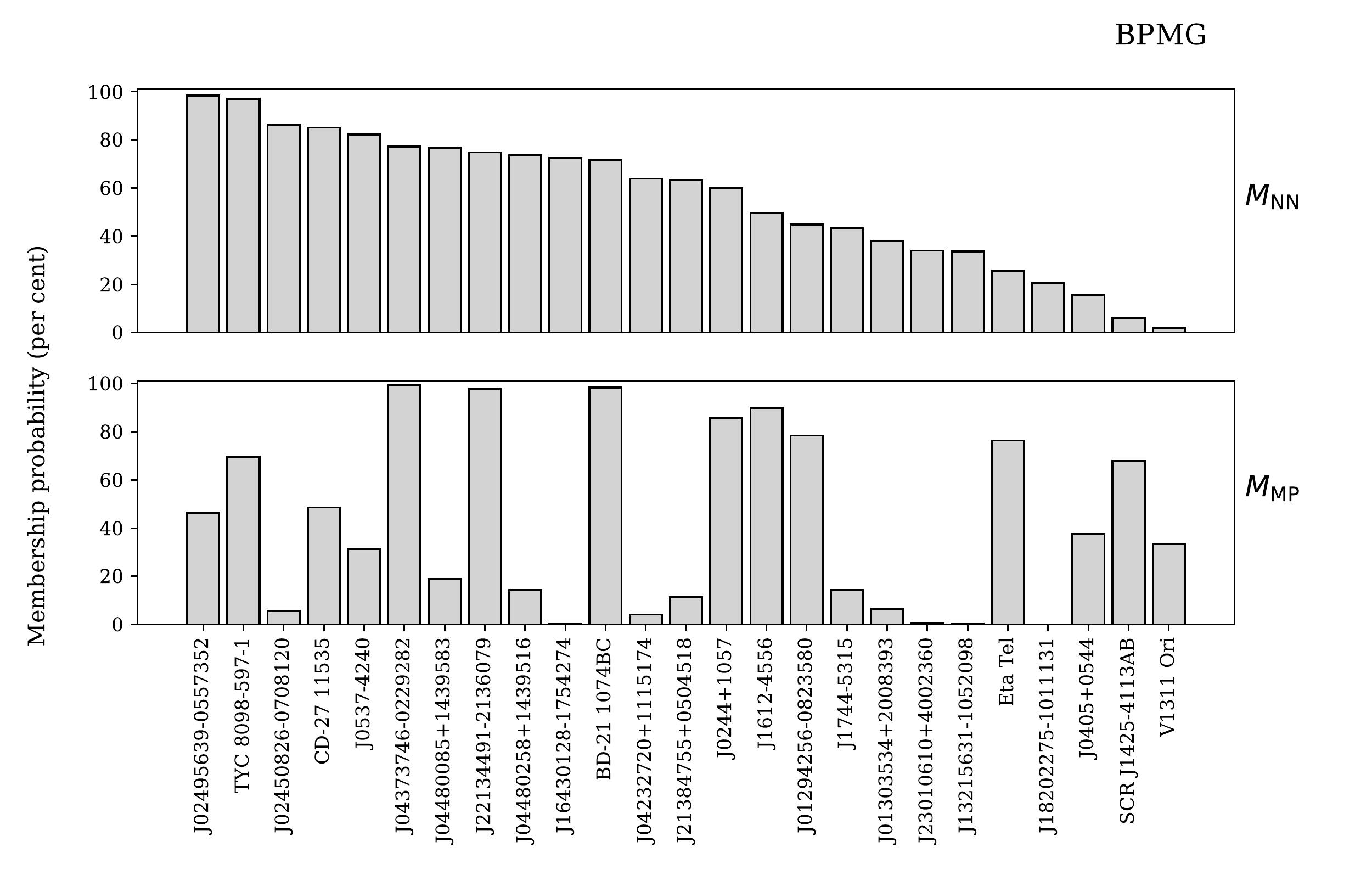}
\caption{Membership probability of claimed BPMG members showing large membership probability difference ($\Delta p>$20 per cent) in the calculation using \mnn\ and \mmp. These stars are presented in Fig.~\ref{fig:bpmg3}.}
\label{fig:bpmg_prob}
\end{figure*}

\subsection{Model development from Stage 1 to 4}

Fig.~\ref{fig:bpmg} illustrates the prototype model (\mo), the revised models (\mnn\ and \mmp), and the pre-final model ($M_{\rm pre-final-(0)}$) for BPMG.
Fig.~\ref{fig:bpmg2} illustrates how a model is finalised in Stage 4.
Model parameters in Stage 1, 2, and 4 are summarized in Table~\ref{fig:mdls}.
As can be seen in Fig.~\ref{fig:bpmg3}, the revised models \mnn\ and \mmp\ are different from each other, which can result in different membership probability for BPMG members.
These stars are displayed in the same figure.
As can be expected, stars match well with \mnn\ have a larger membership probability with \mnn\ (green cross in the figure), while stars agree better with \mmp\ have a larger membership probability with \mmp\ (green circle in the figure).
Membership probabilities of these stars are illustrated in Fig.~\ref{fig:bpmg_prob}.

\subsection{Membership assessment}

In the entire data set, 298 stars have been previously claimed as BPMG members.
According to the categories that described in Section 2.5, there are 113 {\it bona fide}, 2 {\it highly likely}, 30 {\it probable}, 7 {\it possible members}.
These four sets of members are listed in Table~\ref{tab:memb_bpmg}.

On the other hand, 40 stars appear to be not associated with any NYMGs ($p_{\rm field}>$ 80 per cent).
Thirty-three stars likely belong to other NYMGs ($p_{\rm othergroup}>$ 80 per cent), and 11 of these 33 stars have 
already been suggested as members of other NYMGs.
Twenty-two stars claimed as BPMG members in the literature are claimed as members of other NYMGs for the first time in this study.

\begin{table}
\scriptsize
\setlength\tabcolsep{2pt} 
\begin{threeparttable}
\begin{tabular}{*{6}{c}}
\hline
Group & Name & SpT& \ra &\dec  &  p\tnote{a} \\
& & & hh:mm:ss & dd:mm:ss & per cent  \\
 \hline \hline
 \multicolumn{6}{c}{\it bona fide members} \\
     BPMG &                          HD203                  &          F2 &  00:06:50.09 & -23:06:27.1 & 100.0 \\
     BPMG &        2MASS J00172353-6645124 &       M2.5V &  00:17:23.54 & -66:45:12.3 & 100.0 \\ 
     BPMG &        2MASS J00194303+1951117 &        M4.7 &  00:19:43.04 & +19:51:11.7 &  98.5 \\
\multicolumn{6}{c}{\dots} \\
\multicolumn{6}{c}{\dots} \\
\multicolumn{6}{c}{\dots} \\
 \multicolumn{6}{c}{\it highly likely members} \\
    BPMG &        2MASS J04435686+3723033 &          M3 &  04:43:56.87 & +37:23:03.4 &  86.2 \\
    BPMG &        2MASS J06131330-2742054 &        M3.5 &  06:13:13.32 &  -27:42:05.5 &  89.7 \\ 
 \multicolumn{6}{c}{\it probable members} \\
     BPMG &        2MASS J01294256-0823580 &          M7 &  01:29:42.57 & -08:23:58.2 &  99.9 \\
     BPMG &                     J0152+0833              &          M3 &  01:52:57.36 & +08:33:25.9 & 100.0 \\
     BPMG &                     J0216+3043              &          M4 &  02:16:02.49 & +30:43:57.4 & 100.0 \\
      \multicolumn{6}{c}{\dots} \\
\multicolumn{6}{c}{\dots} \\
\multicolumn{6}{c}{\dots} \\
    \multicolumn{6}{c}{\it possible members} \\
     BPMG &        2MASS J01365516-0647379 &     M4V+>L0 &  01:36:55.18 & -06:47:38.0 & 100.0 \\
     BPMG &        2MASS J01373545-0645375 &          G9 &  01:37:35.47 & -06:45:37.5 & 100.0 \\ 
   \multicolumn{6}{c}{\dots} \\
\multicolumn{6}{c}{\dots} \\
\multicolumn{6}{c}{\dots} \\
\hline
\end{tabular}
\begin{tablenotes}
\item[a] Membership probability calculated using \mfin.
\end{tablenotes}
\end{threeparttable}
\caption{Selected BPMG members in 4 membership classes. 
The entire table is available online.
1) {\it Bona fide members} are members having a membership probability of $>$90 per cent, full astrometric parameters, and confirmed youth.
2) {\it Highly likely members} are members having full astrometric parameters and confirmed youth, but with slightly low membership probability (80$-$90 per cent).
3) {\it probable members} are members having membership probability of $>$90 per cent and confirmed youth, but missing \rv\ or \plx.
4) {\it Possible members} are members having a membership probability of $>$90 per cent and full astrometric parameters, but their youth is not confirmed. }
\label{tab:memb_bpmg}
\end{table}

\section{Discussion}

\subsection{Impact of the chosen initial set of members to the finalised models}

One could assume that some biases may be introduced by using the small number of samples in the NYMG discovery paper, and it could make biased results to the finalised models (i.e., stars in the discovery paper may represent a small region of the entire distribution of $XYZ/UVW$ occupied by the entire set of true members).
To resolve this possible issue, we checked distributions of all initial members for each NYMG and confirmed that their distributions are not biased (i.e., for each NYMG, distributions of initial members in XYZ and UVW coincide to those of all claimed members) except for Carina.  
For all other NYMGs excepting for Carina, a slight change in the distribution of initial members would not cause any change in the final NYMG models.  
As a test, we build up ABDor models in two different ways: (1) omission of some initial members (50 per cent)  and (2) initial models with central positions shifted up to 40 per cent  [relative to the 2$\sigma$ length of the major axes (XYZ and UVW) of the model].  
The results show that these differences did not cause any change in the finalised NYMG models. \newline

\begin{figure*}
\includegraphics[width=0.99\textwidth]{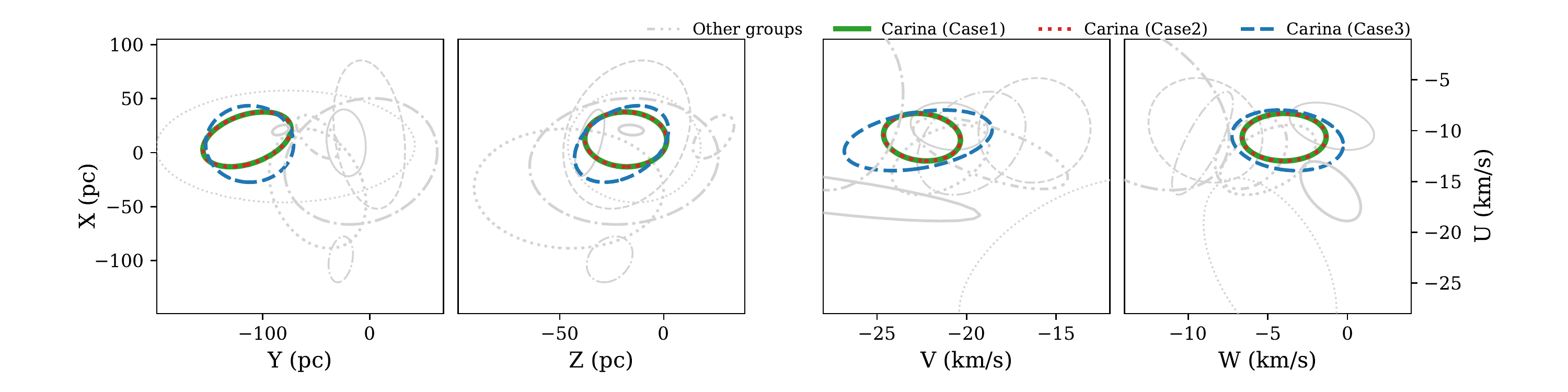}
\caption{The finalised Carina models starting different initial sets in Stage 1.
Models of other groups are displayed with grey colors for comparison.
Models built by using case 1, 2, and 3 are presented with green solid, red dot, and blue dashed lines, respectively.
For more details, see Section 4.1.}
\label{fig:testcarina}
\end{figure*}

A single exception, Carina, has initial members distributed in the half of the entire $X$ distribution of all claimed members.
While all claimed members are distributed in -80$< X$ (pc) $ <$+50, the initial members are distributed in -5$<X$ (pc) $ <$+50. 
We compared results from three cases: (1) using the entire initial members, (2) using a subset of manually picked initial members that are close to the centre positions in XYZ and UVW for the entire set of claimed members, and (3) selecting starting members from the opposite half occupied by initial members.  
Our iterative procedure of ingesting additional members converged and produced very similar results among these three cases (Figure~\ref{fig:testcarina}). 
Case (1) and (2) result in the identical model, and Case (3) produced a finalised model with centre positions of $XYZ$ and $UVW$ agree within 85 per cent.
Up to four members only appear in one of three case results, and it caused a slight change in the extension and orientation of XYZ/UVW ellipsoids.  
One might concern that Case (3) results in a significantly different finalised model compared to the original result.
However, this case is the extreme one presuming all the original initial members are false and starting with the exclusive set of different initial members.
Therefore, we can conclude that our method would not cause any important bias.

\subsection{Mass distribution}

We created a new list of bona fide members for BPMG in Section 3.2 (for other groups, in Appendix B).
Binaries were noted using a binary catalogue by \citet{elb18}.
Utilising these bona fide members, we examined the distribution of spectral types of each NYMG, the distribution of spectral types and the mass function for the entire population of NYMGs.
Stellar masses were interpolated using spectral types and theoretical isochrones \citep{sie00} in the spectral range of B to M6.

\paragraph*{Spectral type distribution}

\begin{figure*}
\includegraphics[width=0.99\textwidth]{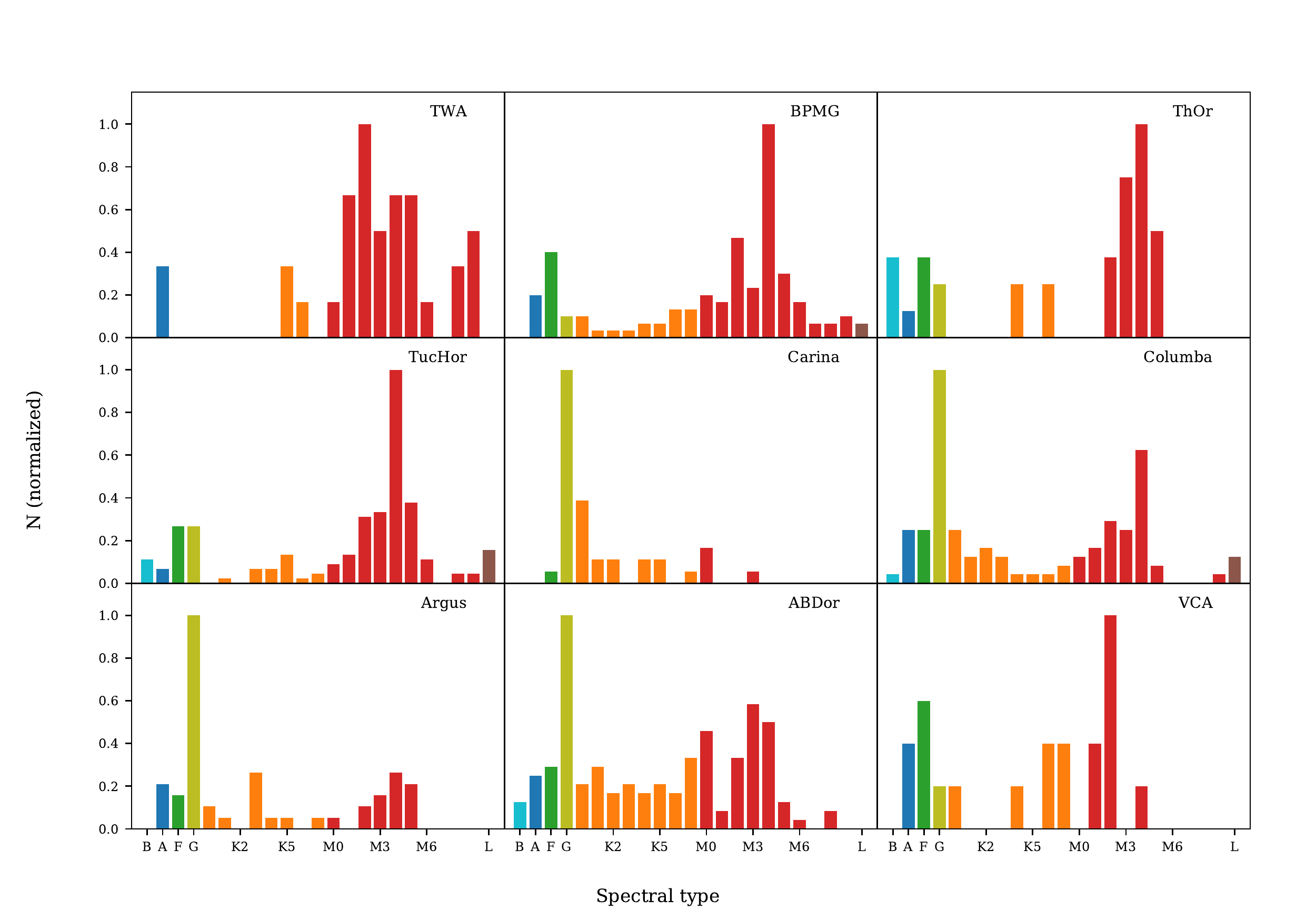}
\caption{The normalised histogram of the spectral type distribution of bona fide members for each NYMG.
Colurs indicate spectral types: cyan, blue, green, yellow, orange, red, and brown correspond to B to L types.}
\label{fig:spt_distrib}
\end{figure*}

Fig.~\ref{fig:spt_distrib} presents spectral type distributions of each NYMG.
Since the majority of NYMG members have masses smaller than 1\msun, we displayed K- and M-types in more details in sub-spectral classes.
Colours in Fig.~\ref{fig:spt_distrib} indicate spectral types: cyan, blue, green, yellow, orange, red, and brown correspond to B to L types.

Each group shows a different spectral type distribution.
TWA mainly consists of M type members, and only five members are earlier than M-type (2 A-type and 3 K-type members).
While BPMG, ThOr, and TucHor members show different distributions of spectral types, they have a common peak at M4 in the histogram.
ThOr seems to have a relatively large fraction of B-, A-, and F-type members compared to other NYMGs.
Additionally, this group seems to be deficient of K-type members.
Four other groups (Carina, Columba, Argus, and ABDor) have a peak at G-type in the histogram.
Carina has relatively many G-type members but lack of late-type members.
Distributions of Columba and ABDor look similar.
They have a larger fraction of K- and M-type members than that of Carina.
However, these two groups also appear to lack low-mass members compared to TWA, BPMG, ThOr, and TucHor.
VCA has a peak at M2 in the histogram, which indicates a lack of later type members as well.

\paragraph*{Mass function}

\begin{figure*}
\includegraphics[width=0.99\textwidth]{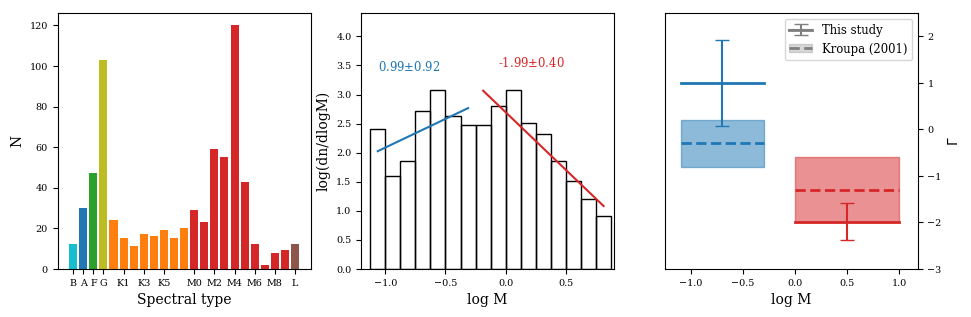}
\caption{Left: spectral type distribution of the entire population of bona fide NYMG members (N=704). Middle: mass function of the members. 
The exponents of power-law [$\xi(\mbox{log}m) \propto m^{\Gamma}$] fits are obtained in two mass regimes: (1) $>$0.5\msun\ and (2) 0.08$-$0.5\msun.
Due to the significant uncertainty of the deduced masses, substellar region (M$<$0.08\msun) is not presented.
Numbers correspond to the $\Gamma$ values.
Right: comparison of the exponent of the power-law for NYMGs and those from \citet{kro01}.
These values agree within the uncertainty of each value.}
\label{fig:mf_all}
\end{figure*}

The middle panel in Fig.~\ref{fig:mf_all} displays the mass function for the entire population of the bona fide NYMG members.
Exponents of power-law [$\xi(\mbox{log}m) \propto m^{\Gamma}$] fits are obtained.
The mass regime is split into (1) $>$0.5 \msun\ and (2) 0.08 $-$ 0.5 \msun.
The lowest mass regime ($<$0.08 \msun) is not displayed due to the significant uncertainty of the deduced mass.
The right panel in Fig.~\ref{fig:mf_all} compares the $\Gamma$ values from this study and those from \citet{kro01}.
In both mass regimes, the $\Gamma$ values of NYMGs and those from \citet{kro01} match within the uncertainty range.
However, if we accept the exact $\Gamma$ value in \citet{kro01}, hundreds of more late-M type NYMG members are expected to be identified.
The deficiency of low mass members is a selection effect due to limitations in member detection methods in the lowest mass regime prior to Gaia DR2.
We expect that Gaia DR2 should enable identification of almost all stellar members of NYMGs within 100 pc of Earth (Lee \& Song in prep.).

\subsection{Kinematic evolution of NYMGs}

We consider the kinematic evolution of NYMGs.
For obtaining a rough idea, we made several assumptions: 
(1) NYMGs are dissolved over time, 
(2) the kinematic dispersion rate is constant over time, 
(3) there is no dynamic perturbation among members,
(4) 3$\times\tilde\sigma_{\rm xyz}$ (3 times the volume averaged size, $\sqrt[3]{\sigma_x \sigma_y \sigma_z}$) is a representative size of NYMGs (in a Gaussian distribution, $\sim$90 per cent of stars are enclosed in this 3$\sigma$ volume), and 
(5) NYMGs with a similar total mass (i.e., having a similar number of members) evolve similarly. \newline

The expansion rate is calculated by the difference of 3$\tilde\sigma_{xyz}$ values over the age difference of the two groups.
We split NYMGs into two classes: less populated (N=30$-$40; TWA, ThOr, and Carina) and populated (N$>$85; BPMG, TucHor, Columba, and ABDor).
Argus (N=45) and VCA (N=19) are excluded in this discussion.
When $\sim$30 Myr old groups are compared to younger groups (Columba vs BPMG, Carina vs TWA, and Carina vs ThOr), the expansion rate is approximately 1$-$2 \kms.
When TucHor and ABDor are compared, the expansion rate is approximately 1 \kms.
However, when ABDor is compared against BPMG, the expansion rate appears to be smaller (0.3$-$0.9 \kms). 
Although we cannot confirm the exact value, we can suggest that the approximate expansion rate of NYMGs would be $\sim$1 \kms.
Virial velocities for NYMGs utilising 3$\tilde\sigma_{xyz}$ and total mass of bona fide members are smaller than 0.15 \kms, which indicate that NYMGs are in the supervirial state.
It is well known that young stellar clusters are often in the supervirial state \citep{gie10, kuh19}.
As noted by \citet{goo06}, the supervirial state can be related to the gas expulsion from the parent molecular cloud.

\subsection{Comparison to BANYAN $\Sigma$}

For assessment of NYMG memberships, several schemes [e.g., BANYAN I, II, $\Sigma$ \citep{mal13,gag14a,gag18b}, LACEwING \citep{rie17}] have been developed and used.
BANYAN series and our scheme have a similar calculation method.
However, mainly due to the difference in models, membership probabilities are different from these methods.
One of the largest difference appears in the Carina model along the y-axis.
For this group, the centre positions in Y from this study and BANYAN $\Sigma$ are -115 pc and -50 pc, respectively.
In this study, the Carina model is developed beginning with the entire initial members and finalised through a self-consistent iterative process.
On the other hand, the Carina model in BANYAN $\Sigma$ is created using seven nearest members.
As we demonstrated in Section 4.1., if the BANYAN Carina model was developed via an iterative process, this model and our finalised Carina model would be similar.
In addition, when we search for low-mass Carina members from Gaia DR2 (Lee \& Song in prep.), a significant number of distant Carina members ($>$100 pc and over-luminous on a colour-magnitude diagram) are identified, which is indicative of the nature of Carina group being large and distant. \newline

In this section, we compare membership probability calculation from BANYAN $\Sigma$ \citep{gag18b} and our scheme (BAMG, paper 1 and this paper) focusing on objects of particular interest (Table~\ref{tab:comp_prob}, ~\ref{tab:comp_prob2}, and ~\ref{tab:planet}).
Membership probabilities from BAMG and BANYAN $\Sigma$ for the entire data $D$ are included in Table~\ref{tab:entiredata} (online version).

\begin{table}
\centering
\small %\tiny%\footnotesize %\small%\scriptsize
\setlength\tabcolsep{2pt} 
\begin{threeparttable}
 \begin{tabular}{*{4}{c}}
 \toprule
 Name & SpT  & \multicolumn{2}{c}{$p_{\rm max}$(group)}  \\ 
 \cmidrule{3-4} 
 & &BAMG (this study)  & BANYAN $\Sigma$ \\
 \hline
 TWA 6                                       & M0Ve     & 100 (TWA)   & 100 (Field) \\
 TWA 2	                                 & M2Ve	 & 100 (TWA)  & 96 (Field) \\ 
TWA 3B	                                 & M4Ve	& 100 (TWA)   &	89 (Field) \\
TWA 16	                                  & M1Ve	& 94 (Field)	& 96 (TWA) \\
HD 2885	                                 &A2V	&99 (TucHor)	&87 (Field)\\
HD 12894	                                  &F2V	&100 (TucHor)	&100 (Field)\\
2MASS J06085283-2753583   &M9+L0	 &100 (Columba)	&100 (Field)\\
HIP 63742                           	&       K1	       &98 (ABDor)       &	71 (Field)\\
\bottomrule
\end{tabular}
\begin{tablenotes}
\item[\dag] Input parameters necessary to calculate membership probabilities are listed in Table~\ref{tab:entiredata}, so we do not repeat them here. For a given star, BAMG and BANYAN $\Sigma$ calculate membership probabilities to belong a certain NYMG for all major MGs ($N$=9) including the field star population.  In this table, we list the membership probability of the designated NYMG listed in the literature calculated based on BAMG and BANYAN $\Sigma$. 
\end{tablenotes}
\end{threeparttable}
\caption{Stars showing significantly different membership probabilities from our scheme (BAMG) and BANYAN $\Sigma$. Only a subset of members is presented.\tnote{\dag}
For other members not listed in this table, see Table~\ref{tab:entiredata}.} 
\label{tab:comp_prob}
\end{table}

\begin{table}
\centering
\small %\tiny%\footnotesize %\small%\scriptsize
\setlength\tabcolsep{2pt} 
\begin{threeparttable}
 \begin{tabular}{*{4}{c}}
 \toprule
 Name & SpT  & \multicolumn{2}{c}{$p$ (group)}  \\ 
 \cmidrule{3-4} 
 & &BAMG (this study)  & BANYAN $\Sigma$ \\
 \hline
 HD 2884  & B9V & 100 (TucHor) & 59 (TucHor) \\
 HD 2885  & A2V & 100 (TucHor) & 87 (Field) \\
 \hline
HIP 12635 & K3.5 & 100 (ABDor) & 97 (ABDor) \\
HIP 12638 & G5 & 100 (Field) & 100 (Field) \\
 \hline
  ome Aur A & A1 & 35, 34 (BPMG, ABDor) & 99 (Columba) \\
 ome Aur B & F9 & 97 (ABDor) & 77 (Columba) \\
 \hline
 HD 64982 & G0V & 61, 24 (Columba, ABDor) & 55, 27 (Field, ABDor) \\
 HD 64982B & K & 98 (Field) & 97 (Field) \\
 \hline
 HD 83096 & F2V & 62 (Carina) & 100 (Carina) \\
 HIP 46720B & G9V & 52 (Argus) & 92 (Field) \\
 \hline
 $\eta$ Tel & A0 & 96 (BPMG) & 77 (BPMG) \\
 HR 7329B & M7/8V & 96 (BPMG) & 97 (BPMG) \\
 \bottomrule
\end{tabular}
%\begin{tablenotes}
%\end{tablenotes}
\end{threeparttable}
\caption{Binary members showing inconsistency between the members in terms of membership probability.
Binary systems showing inconsistent membership status with either BAMG or BANYAN $\Sigma$ are presented.
Horizontal lines separate different systems.}
\label{tab:comp_prob2}
\end{table}

\subsubsection{Some well-known members}

Table~\ref{tab:comp_prob} lists some stars showing a different membership status based on BAMG and BANYAN $\Sigma$. \newline

TWA 2, 3, and 6 were discovered as members of TWA \citep{web99}, and TWA 2 and 3 are the first discovery members of a co-moving group of 5 nearby T Tauri stars \citep{kas97}.
While TWA 3A has a large membership probability (nearly 100 per cent) from both BAMG and BANYAN $\Sigma$, TWA 3B has a large field star probability from BANYAN $\Sigma$ as well as TWA 2 and 6. \newline

TWA 16 has a large TWA membership probability with BANYAN $\Sigma$ (96 per cent) while it is likely to be a field star from BAMG (94 per cent).
\citet{scn12a} listed this star as a bona fide member while \citet{duc14} claimed this star as a possible member. \newline

HD 2885 and HD 12894 are initial TucHor members \citep{zuc00, tor00}.
BANYAN $\Sigma$ assesses these members as field stars ($>$85 per cent).
While these two stars are off in UVW [HD 2885 has small $U$ (-15 \kms) and $V$ (-26 \kms) values and HD 12894 has a large $V$ (-18 \kms) value], they are located nearly at the centre of the spatial TucHor model (XYZ) of our scheme. \newline

2MASS J06085283-2753583 was suggested as a BPMG member from \citet{ric10}, then suggested as a peripheral Columba member from \citet{gag14a}.  
\citet{gag14a} mentioned that 2MASS J06085283-2753583 shows signs of youth from various studies [signs of low-gravity \citep{all13} and strength of Li feature \citep{kir08}]. 
This object is evaluated as a Columba member from BAMG (100 per cent) while it can be a field star from BANYAN $\Sigma$ (100 per cent). \newline

One initial ABDor member HIP 63742 is assessed as an ABDor member from BAMG (98 per cent) while a field star from BANYAN $\Sigma$ calculation (71 per cent). \newline

\subsubsection{Wide binaries}

For stars in binary systems, if a binary belongs to a NYMG, then both members should belong to the same NYMG.
In this section, we present binary members having inconsistent membership probabilities in a system with either BAMG or BANYAN $\Sigma$ (Table~\ref{tab:comp_prob2}). \newline

HD 2884 and HD 2885 are initial TucHor members.
Both HD 2884 and HD 2885 are binaries (WDS J00315-6257AB and WDS J00315-6257CD, respectively).
BAMG consistently assesses both stars as members of TucHor while BANYAN $\Sigma$ assesses HD 2884 and HD 2885 as a marginal TucHor member and a field star, respectively. \newline

HIP 12635 and HIP 12638 are initial ABDor members \citep{zuc04a}.
In both BAMG and BANYAN $\Sigma$ calculation, HIP 12635 is assessed as an ABDor member while HIP 12638 is assessed as a field star.
Proper motions of these two stars agree within 10 per cent difference.
Radial velocities of two stars are -3.3 \kms\ and -4.7 \kms, respectively.
The main reason causing the different membership status from both methods is the distance.
Distances to HIP 12635 and HIP 12638 are 50 pc and 71 pc, respectively. \newline

Ome Aur A and ome Aur B are claimed as Columba members by \citet{zuc11} and \citet{ell16}.
BANYAN $\Sigma$ assessed both as Columba members, with a marginal membership for ome Aur B (77 per cent).
On the other hand, with BAMG, ome Aur B is assessed as an ABDor member while ome Aur A is assessed as an ambiguous member of ABDor or BPMG. \newline

HD 64982 and HD 64982B are claimed as ABDor members by \citet{tor08} and \citet{ell16}.
While BAMG and BANYAN $\Sigma$ assessed HD 64982 as a marginal Columba member and a field star, respectively, both calculations provide similar low ABDor membership probabilities ($\sim$25 per cent).
In both methods, HD 64982B is assessed as a field star.
The main factor of the inconsistent membership status with a given method might be lacking RV for HD 64982B. \newline

HD 83096 and HIP 46720B (HD 83096B) are initial Carina members \citep{tor08}.
With BAMG, HD 83096 and HIP 46720B are assessed as a marginal Carina member and a marginal Argus member, respectively. 
With BANYAN $\Sigma$, while HD 83096 is assessed as a Carina member, HIP 46720B is evaluated as a field star. \newline

$\eta$ Tel and HR 7329B are initial BPMG members \citep{zuc01b}.
With BAMG, both stars are assessed as BPMG members.
With BANYAN $\Sigma$, HR 7329B has a large BPMG membership probability, while $\eta$ Tel has a marginal membership probability.

\begin{table*}
\small
\setlength\tabcolsep{2pt} 
\begin{threeparttable}
\begin{tabular}{p{4cm}p{1cm}p{3cm}p{3cm}p{6cm}}%{*{5}{c}}
\toprule
Name & SpT  & \multicolumn{2}{c}{$p$ (group)}  & Planet name \\
\cmidrule{3-4} 
& & BAMG (this study) & BANYAN $\Sigma$ \\
 \hline 
TYC 9486-927-1	& M1Ve	& 100 (BPMG)	& 100 (Field)	& 2M J2126-81 b \\
2MASS J22501512+2325342A	& M3	 & 99 (ABDor) & 55 (ABDor), 45 (Field) &	2MASS J2250+2325 b \\
AB Pic	& K1Ve	& 100 (Columba) & 99 (Carina) &	AB Pic b \\
CD-52 381	& K2Ve	& 99 (Columba) & 46 (Columba), 54 (Field) &	GSC 08047-00232 b \\
HIP 1134	& F5	 & 54 (BPMG), 41 (ABDor) & 100 (Field)&	HD 984 b \\
HIP 114189	& A5	& 87 (BPMG), 12 (ABDor) &49 (Columba), 51 (Field)	& HR 8799 b, HR 8799 c, HR 8799 d, HR 8799 e\\
HD 222439	&B9	& 86 (BPMG) &	92 (Field), 8 (Columba) &	$\kappa$ And b \\
\bottomrule
\end{tabular}
%\begin{tablenotes}
%\end{tablenotes}
\end{threeparttable}
\caption{Planet host moving group members showing significantly different membership probabilities from our scheme (BAMG) and BANYAN $\Sigma$. }
\label{tab:planet}
\end{table*}

\subsubsection{planet host stars}

NYMGs are important for studying exoplanets.
They are prime targets for exoplanet imaging. 
Age of host stars can be constrained by the group age, which can constrain the mass of planets.
The online exoplanet archive (exoplanet.eu) presents 98 planetary systems detected by the imaging method.
Twenty-two systems are cross-matched with the entire set of claimed NYMG members.

Table~\ref{tab:planet} presents a subset of these systems (7 systems) showing a significantly large membership probability difference from our scheme and BANYAN $\Sigma$.

Three stars have large field star probabilities from BANYAN $\Sigma$.
Two of them (TYC 9486-927-1 and HD 222439) have large BPMG membership probabilities ($>$80 per cent) from BAMG. 
The remaining star (HIP 1134) has a marginal membership probability belonging to either BPMG or ABDor.
Revised membership status can change the age of the host star.
Therefore, the mass of the planets can be changed.
For these three stars, the changed membership status can significantly affect the characteristics of planets because of the large age difference (field population vs BPMG).

On the other hand, 2MASS J22501512+2325342, AB Pic, and CD-52 381 would receive relatively smaller impacts caused by a different membership status based on the two methods.
For example, 2MASS J22501512+2325342 is assessed as a Columba member with BAMG, while BANYAN $\Sigma$ assesses this star as a marginal Columba member.
Even though AB Pic is assessed as a Columba member with BAMG but as a Carina member with BANYAN $\Sigma$, these two NYMGs have a similar age, which will not change the derived planetary mass.

The remaining star HIP114189 (HR 8799) is a popular multiple planet host star (at least four planets).  
This star is known as a Columba member \citep{mar10}, and BANYAN $\Sigma$ returns a marginal Columba membership probability (49 per cent).
Our calculation suggests this star as a BPMG member (87 per cent).
If that is true, the masses of the exoplanets (b, c, d, e) are reduced to the factor of 2 [based on Figure 3 in \citet{mar10}].

\section{Conclusion and summary}

Since the discoveries of NYMGs, many members of different quality were suggested as additional members.
These members can be utilised to build up NYMG models and identify new members.
Unscreened acceptance of all these suggested members can lead to distorted NYMG models, and hence bias in identification of new members.
Therefore, a careful investigation and selection of members and model constructions are required.\newline

In this study, NYMG models were established through 4 stages.
Initially, the prototype models were created utilising the initial members.
Outliers were rejected based on the nearest neighbour distance.
In Stage 2, the prototype models were revised by incorporating appropriate additional members evaluated by two methods (nearest neighbour distance and membership probability).
In Stage 3 and 4, the models were finalised utilising members having large membership probabilities regardless of the model revision method. 
We tested different sets of samples in Stage 1 to resolve a possible bias due to initial list dependency.
The results show that this effect is not significant. \newline

For the entire set of NYMG members (N=1913), about 35 per cent of members are confirmed as bona fide members while $\sim$13 per cent of them are evaluated as field stars ($p_{\rm field}>$80 per cent).
As the main results of this study, we present lists of bona fide members and model parameters of NYMGs.
Distributions of spectral types and the mass functions of the entire NYMG members show a deficiency of M-type members.
This is a selection effect in the lowest mass regime prior to Gaia DR2.
We expect that Gaia DR2 should enable the identification of almost all stellar members of the NYMGs within 100 pc of Earth (Lee et al. in prep.). \newline

While we obtain consistent NYMG models via our iterative procedure, there are differences among NYMG models and calculation schemes among NYMG calculation methods (e.g., BANYAN series), which can significantly affect membership assessments.
The different membership assessment can have significant impacts on the relevant studies.
For example, seven planet-host stars have different membership status from our scheme and BANYAN $\Sigma$, which can change the age and the mass of the planets.
In the case of HR 8799, by changing membership from Columba ($\sim$30Myr) to BPMG ($\sim$15Myr), the masses of planets can be reduced by the factor of 2.
In the future, we need to resolve this issue by developing a unified analysis of young nearby stars and NYMGs.

%%%%%%%%%%%%%%%%%%%%%%%%%%%%%%%%%%%%%%%%%%%%%%%%%%

\section*{Acknowledgements}
We acknowledge the important and critical review by Ben Zuckerman that improved the manuscript significantly.
We also thank the anonymous referee for valuable comments and suggestions that helped to significantly increate the quality of this work.

%%%%%%%%%%%%%%%%%%%%%%%%%%%%%%%%%%%%%%%%%%%%%%%%%%



%\bibliographystyle{mnras}


\section{SUPPORTING INFORMATION}
Additional Supporting Information may be found in the online version of this article. \newline
Appendix A: Additional NYMGs included in this study \newline
Appendix B: Results for the other 8 NYMGs: kinematic NYMG models and lists of members \newline

%%%%%%%%%%%%%%%%%%%%%%%%%%%%%%%%%%%%%%%%%%%%%%%%%%

%%%%%%%%%%%%%%%%%%%% APPENDICES %%%%%%%%%%%%%%%%%%
\clearpage
\appendix

\section{Additional NYMGs included in this study}

In this study, we considered NYMGs whose mean distances and ages are less than 100 pc and 200 Myr, respectively.
Moving groups not included have ages that are less decisive due to the lack of data and the reddening effect.
In addition, membership identification of these groups is also far from being complete.
Nine NYMGs listed in Table~\ref{tab:groups} are considered in this study.

\paragraph*{The TW Hydrae association (TWA)}
\citet{kas97} discovered 5 co-moving X-ray bright stars around TW Hydrae.
\citet{web99} identified $\sim$10 more co-moving members and identified this group of stars as the TW Hydrae association (TWA), which is the firstly identified NYMG.
Classical papers about NYMGs \citep{zuc04b, tor08} characterised the TWA with the age of 8 Myr and the mean distance of 50 pc.
Due to its young age and proximity, members of TWA have been intensively identified ($\sim$100 members from studies listed in Table~\ref{tab:groups}).

\paragraph*{The $\beta$ Pictoris moving group (BPMG)}
Discovered with the $\sim$20 co-moving members by \citet{zuc01b}, the BPMG is the secondly identified NYMG.
The age was suggested as 12 to 24 Myr \citep{zuc01b, bel15}.
Due to the young age and the closest mean distance (35 pc), members of BPMG have been intensively searched.
Approximately 290 members have been claimed from studies listed in Table~\ref{tab:groups}.

\paragraph*{The 32 Ori group (ThOr)}
\citet{mam07} identified a dozen of young co-moving stars around 32 Ori.
The relatively recently discovered group, ThOr, has the mean distance and the age of 90 pc and $\sim$20 Myr, respectively \citep{bel17}.
Fifty-five members have been identified from studies listed in Table~\ref{tab:groups}.

\paragraph*{The Tucana/Horologium association (TucHor)}
\citet{zuc00} discovered the Tucana association consists of $\sim$20 co-moving star systems.
At the similar time, \citet{tor00} discovered the Horologium association consists of $\sim$15 young co-moving stars.
\citet{zuc01c} proposed that these two groups constitute a single widespread stellar group, and called the Tucana/Horologium association.
The estimated age and the mean distance of the group are 30$-$45 Myr \citep{zuc01d, bel15} and 50 pc, respectively.
Approximately 400 members have been identified by studies listed in Table~\ref{tab:groups}.

\paragraph*{The Carina association (Carina)}
\citet{tor08} discovered the Carina association having $\sim$20 co-moving members.
The age and mean distance of this group is 30$-$45 Myr \citep{tor08, bel15} and $\sim$90 pc, respectively.
About 190 additional members have been identified by studies listed in Table~\ref{tab:groups}.

\paragraph*{The Columba association (Columba)}
\citet{tor08} discovered the Columba association consists of $\sim$40 co-moving members.
Despite the age and mean distance of this group is similar to those of the Carina [30$-$42 Myr \citep{tor08, bel15} and $\sim$80 pc], this group occupies a slightly different position in XYZ and UVW.
The number of additionally identified members is $\sim$270 from studies listed in Table~\ref{tab:groups}.

\paragraph*{The Argus association (Argus)}
\citet{tor03b} suggested a sparse Argus association, which might be physically connected to IC 2391 open cluster.
\citet{tor08} suggested that Argus consists of spatially compact IC 2391 members and sparse field members. 
Recently, \citet{mam15} and \citet{gag18a} declined the existence of Argus based on the arguments on \citet{bel15}.
\citet{bel15} could not find a distinctive isochron age for Argus, and concluded that the Argus might be highly contaminated or not be a single coeval moving group.
However, this can be caused by the contamination of the input data (i.e., including many non-members).
\citet{rie17} also have doubts about the existence of the Argus because they could not recover a large portion of the Argus members using their NYMG member identification tool LACEwING.
\citet{zuc19} states that Argus is a true association based on the analysis of kinematics and age of the Argus members.
Argus occupies distinct positions relative to the other NYMGs in UVW, and the members also show ambiguous signs of youth.
Therefore, we include Argus in this study.

Argus has the mean distance of $\sim$100 pc, and the age is slightly older than TucHor and Columba [$\sim$40 Myr, \citet{tor08}].
Among members listed in the discovery paper \citep{tor08}, we considered only their ``field'' members ($\sim$20 members) since most IC2391 members are located at 120$-$160 pc, which is beyond our distance limit.
Approximately 150 members have been identified from studies listed in Table~\ref{tab:groups}.

\paragraph*{AB Doradus moving group (ABDor)}
\citet{zuc04a} discovered a slightly old NYMG [50$-$150 Myr; \citet{zuc04b, bel15}], having large $UVW$ values relative to the other NYMGs.
More than 400 members have been identified from studies listed in Table~\ref{tab:groups}.

\paragraph*{Volans-Carina group (VCA)}
\citet{oh17} discovered a group of comoving 19 stars with Gaia DR1.
The mean distance and the age of the group are 85 pc and 90 Myr, respectively \citep{gag18d}.
\citet{gag18d} suggested $\sim$45 additional members with Gaia DR2.

\section{Results for the other 8 NYMGs: kinematic NYMG models and lists of members}

As was done for BPMG, models are created and finalised through Stage 1 to 4.
For each NYMG, we compared \mo, \mnn, \mmp, and $M\rm_{pre-final-(0)}$ in a relevant figure (e.g., Fig. B1).
Table~\ref{tab:mdls} summarises the model parameters, and members and membership probability with the final model are listed in a relevant table (e.g., Table B1).
Table~\ref{tab:entiredata} lists membership probability of the entire claimed members.

\subsection{TWA}

\begin{figure*}
\includegraphics[width=0.9\textwidth]{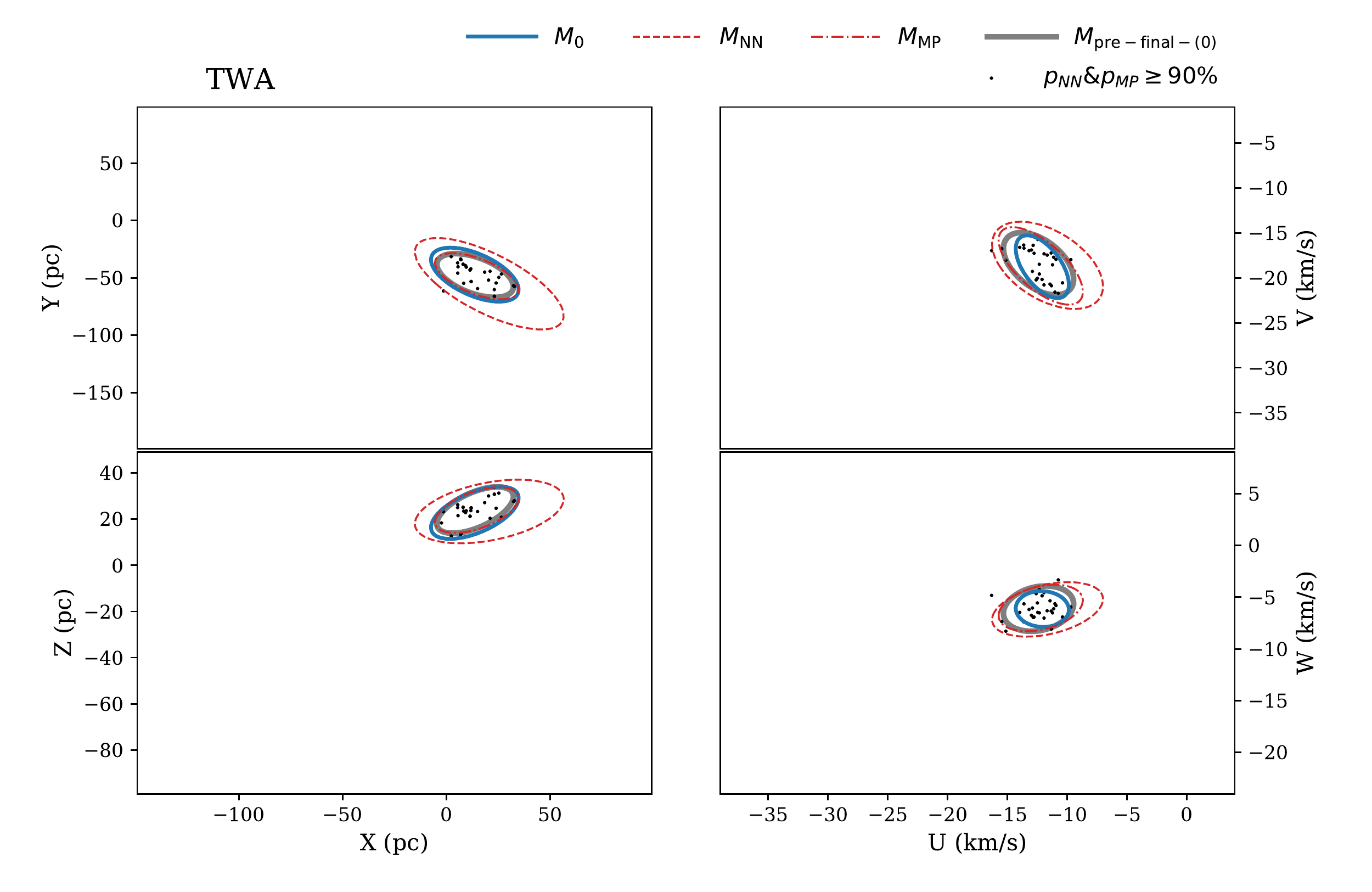}
\caption{Same to Fig.~\ref{fig:bpmg}, but for TWA.}
\label{fig:twa}
\end{figure*}

\begin{table}
\scriptsize
\setlength\tabcolsep{3pt} 
\begin{threeparttable}
\begin{tabular}{*{6}{c}}
\hline
Group & Name & SpT& \ra &\dec  &  p\tnote{a} \\
& & & hh:mm:ss & dd:mm:ss & per cent  \\
 \hline \hline
 \multicolumn{6}{c}{\it bona fide members} \\
       TWA &               SCR J1012-3124AB &        M4.0 &  10:12:09.13 &  -31:24:45.2 & 100.0 \\ 
       TWA &                           TWA6 &        M0Ve &  10:18:28.70 &  -31:50:02.8 & 100.0 \\ 
       TWA &                           TWA7 &        M2Ve &  10:42:30.10 &  -33:40:16.2 &  99.9 \\ 
\multicolumn{6}{c}{\dots} \\
\multicolumn{6}{c}{\dots} \\
\multicolumn{6}{c}{\dots} \\
 \multicolumn{6}{c}{\it probable members} \\
        TWA &        2MASS J10284580-2830374 &          M5 &  10:28:45.80 &  -28:30:37.5 & 100.0 \\
       TWA &        2MASS J10585054-2346206 &        M3.8 &  10:58:50.54 & -23:46:20.6 & 100.0 \\ 
       TWA &        2MASS J11020983-3430355 &        M8.5 &  11:02:09.84 & -34:30:35.6 & 100.0 \\ 
\multicolumn{6}{c}{\dots} \\
\multicolumn{6}{c}{\dots} \\
\multicolumn{6}{c}{\dots} \\
\hline
\end{tabular}
\begin{tablenotes}
\item[a] Membership probability using \mfin.
\end{tablenotes}
\end{threeparttable}
\caption{Selected TWA members in 2 membership classes. 
The description of classes are same as Table~\ref{tab:memb_bpmg}.
The entire table is available online.
}
\label{tab:memb_twa}
\end{table}

In the entire data set, 97 stars have been previously claimed as TWA members.
According to the membership classes that described in Section 2.5., there are 30 {\it bona fide} and 15 {\it probable members}.
These 2 sets of members are listed in Table~\ref{tab:memb_twa}.

On the other hand, 33 claimed members appear to be not associated with any NYMGs ($p_{\rm field}>$ 80 per cent).
Four stars likely belong to other NYMGs ($p_{\rm other group}>$ 80 per cent), and 3 of these 4 stars have already suggested as members of other NYMGs.
One star claimed as a TWA member in the literature is claimed as a member of other NYMGs for the first time in this study.

\subsection{ThOr}

\begin{figure*}
\includegraphics[width=0.9\textwidth]{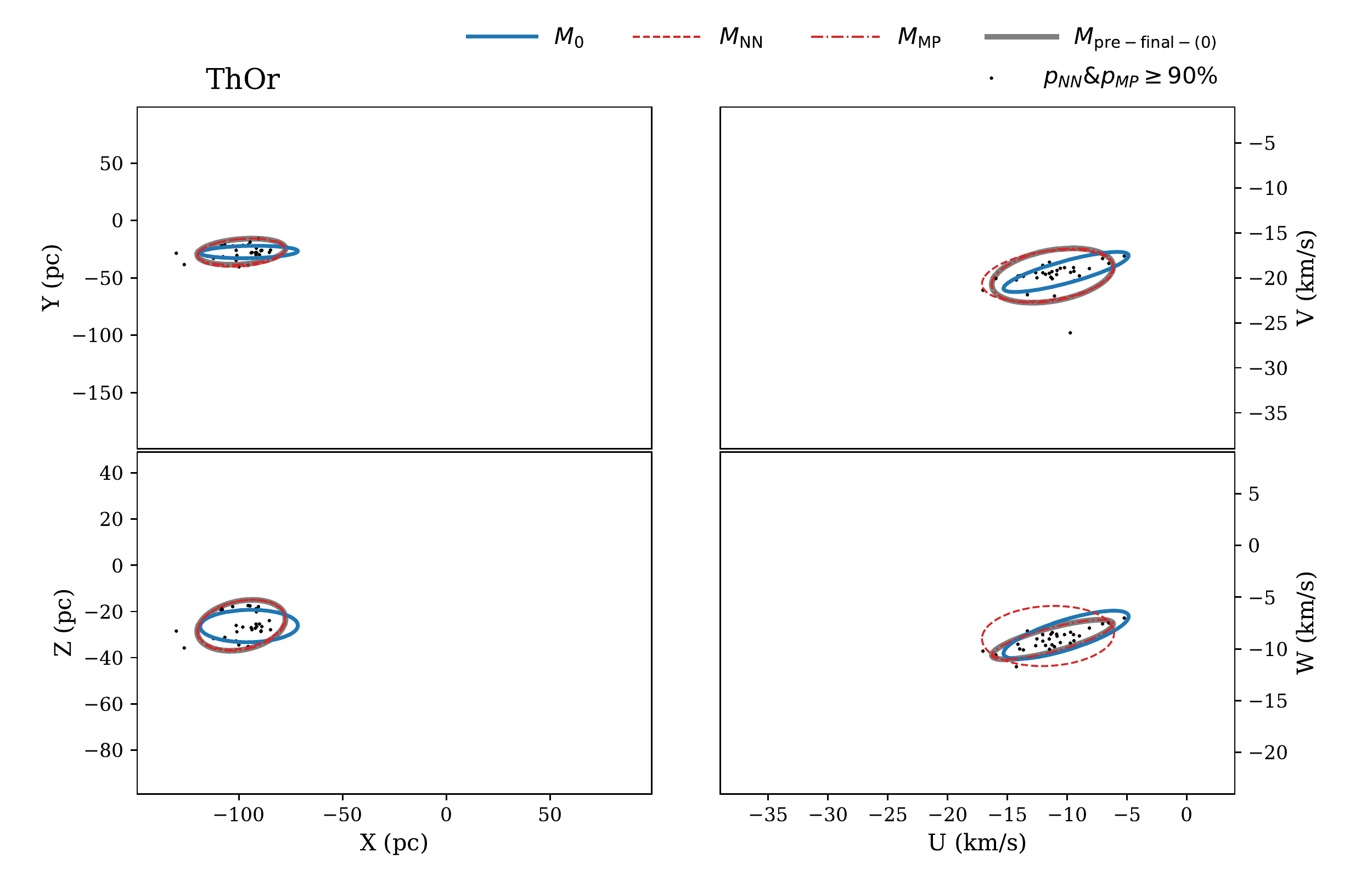}
\caption{Same to Fig.~\ref{fig:twa} but for ThOr.}
\label{fig:thor}
\end{figure*}

\begin{table}
\scriptsize
\setlength\tabcolsep{3pt} 
\begin{threeparttable}
\begin{tabular}{*{6}{c}}
\hline
Group & Name & SpT& \ra &\dec  &  p\tnote{a} \\
& & & hh:mm:ss & dd:mm:ss & per cent  \\
 \hline \hline
 \multicolumn{6}{c}{\it bona fide members} \\
     ThOr &        2MASS J05132631+1057439 &          M3 &  05:13:26.31 &  +10:57:44.0 & 100.0\\
      ThOr &        2MASS J05192941+1038081 &          M3 &  05:19:29.42 &  +10:38:08.2 & 100.0\\ 
      ThOr &        2MASS J05194398+0535021 &          M3 &  05:19:43.99 &  +05:35:02.1 & 100.0 \\
\multicolumn{6}{c}{\dots} \\
\multicolumn{6}{c}{\dots} \\
\multicolumn{6}{c}{\dots} \\
%\hline
 \multicolumn{6}{c}{\it probable members} \\
      ThOr &                     J0520+0511 &          K0 &  05:20:17.95 &  +05:11:52.1 & 100.0\\ 
      ThOr &                     J0527+0626 &          M6 &  05:27:28.05 &  +06:26:43.8 & 100.0 \\
\multicolumn{6}{c}{\dots} \\
\multicolumn{6}{c}{\dots} \\
\multicolumn{6}{c}{\dots} \\
\hline
\end{tabular}
\begin{tablenotes}
\item[a] Membership probability using \mfin.
\end{tablenotes}
\end{threeparttable}
\caption{Selected ThOr members in 2 membership classes. 
The description of classes are same as Table~\ref{tab:memb_bpmg}.
The entire table is available online. }
\label{tab:memb_thor}
\end{table}

In the entire data set, 55 stars have been claimed as ThOr members.
There are 31 {\it bona fide} and 5 {\it probable members}.
These 2 sets of members are listed in Table~\ref{tab:memb_thor}.

On the other hand, 2 stars appear to be not associated with any NYMGs ($p_{\rm field}>$ 80 per cent).
Six stars claimed as ThOr members in the literature are claimed as members of other NYMGs for the first time in this study.

\subsection{TucHor}

\begin{figure*}
\includegraphics[width=0.9\textwidth]{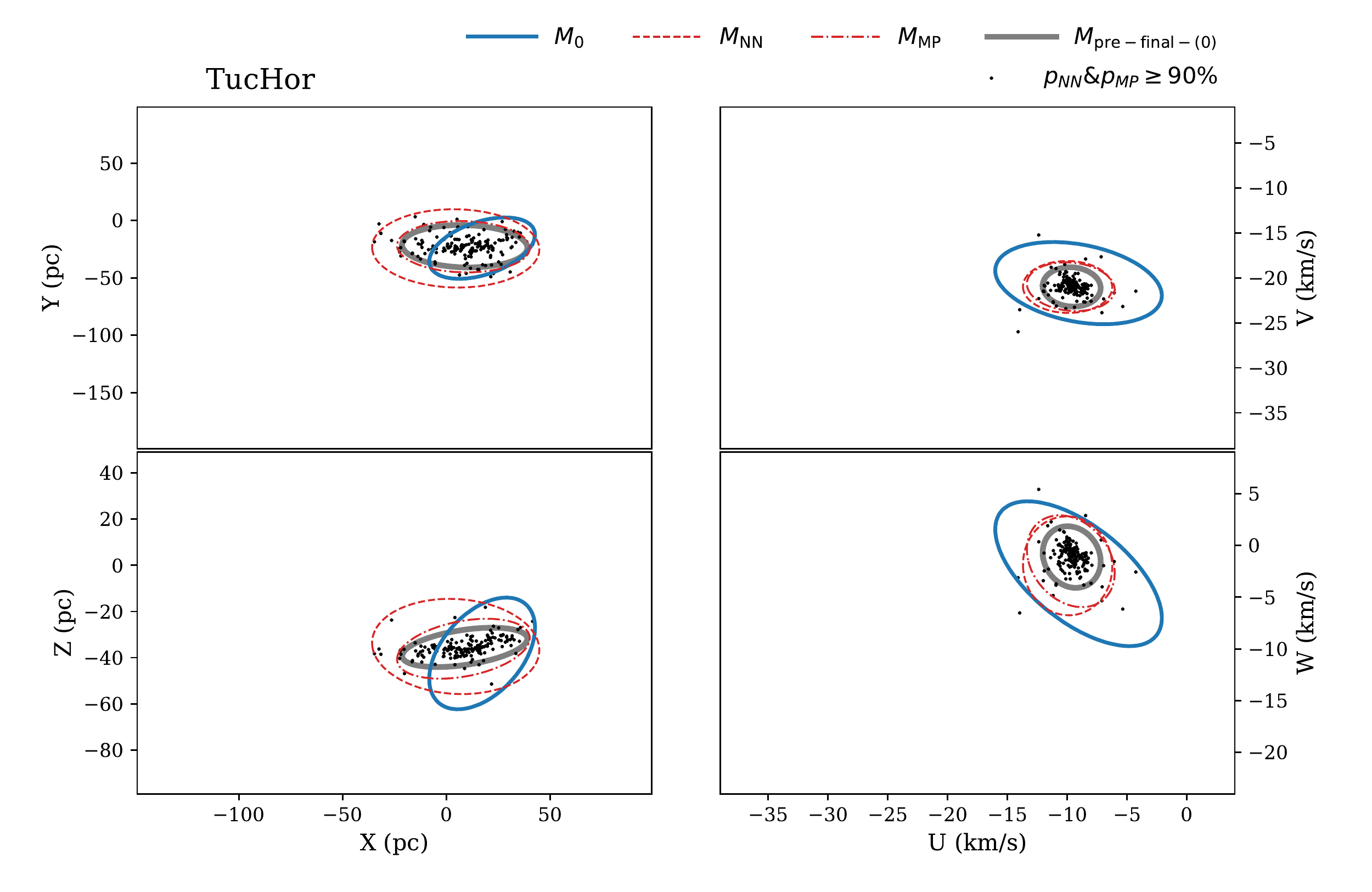}
\caption{Same to Fig.~\ref{fig:twa} but for TucHor.}
\label{fig:tuchor}
\end{figure*}

\begin{table}
\scriptsize
\setlength\tabcolsep{3pt} 
\begin{threeparttable}
\begin{tabular}{*{6}{c}}
\hline
Group & Name & SpT& \ra &\dec  &  p\tnote{a} \\
& & & hh:mm:ss & dd:mm:ss & per cent  \\
 \hline \hline
 \multicolumn{6}{c}{\it bona fide members} \\
    TucHor &                         HD 105 &         G0V &  00:05:52.54 & -41:45:11.0 & 100.0\\
    TucHor &        2MASS J00125703-7952073 &        M2.9 &  00:12:57.05 & -79:52:07.3 & 100.0 \\
    TucHor &                         HD 987 &         G6V &  00:13:53.01 & -74:41:17.9 & 100.0 \\ 
\multicolumn{6}{c}{\dots} \\
\multicolumn{6}{c}{\dots} \\
\multicolumn{6}{c}{\dots} \\
\multicolumn{6}{c}{\it highly likely members} \\
    TucHor &                     J0015-2946 &          M4 &  00:15:36.71 &  -29:46:00.5 &  85.6 \\ 
 \multicolumn{6}{c}{\it probable members} \\
    TucHor &        2MASS J00041589-8747254 &        M5.7 &  00:04:15.84 & -87:47:25.4 & 98.6  \\
    TucHor &        2MASS J00065794-6436542 &         M9: &  00:06:57.93 & -64:36:54.2 & 100.0 \\
    TucHor &        2MASS J00182834-6703130 &        M9.6 &  00:18:28.33 & -67:03:13.0 & 100.0 \\
\multicolumn{6}{c}{\dots} \\
\multicolumn{6}{c}{\dots} \\
\multicolumn{6}{c}{\dots} \\
\multicolumn{6}{c}{\it possible members} \\
    TucHor &                       HD 10269 &         F5V &  01:39:07.62 & -56:25:45.8 & 100.0 \\ 
\hline
\end{tabular}
\begin{tablenotes}
\item[a] Membership probability using \mfin.
\end{tablenotes}
\end{threeparttable}
\caption{Selected TucHor members in 4 membership classes. 
The description of classes are same as Table~\ref{tab:memb_bpmg}.
The entire table is available online.}
\label{tab:memb_tuchor}
\end{table}

In the entire data set, 408 stars have been claimed as TucHor members.
There are 160 {\it bona fide}, 1 {\it highly likely}, 68 {\it probable}, and 1 {\it possible members}.
These 4 sets of members are listed in Table~\ref{tab:memb_tuchor}.

On the other hand, 29 stars appear to be not associated with any NYMGs ($p_{\rm field}>$ 80 per cent).
Eighty-six stars likely belong to other NYMGs ($p_{\rm other group}>$ 80 per cent), and 38 of these 86 stars have already suggested as members of other NYMGs.
Forty-five stars claimed as TucHor members in the literature are claimed as members of other NYMGs for the first time in this study.

\subsection{Carina}

\begin{figure*}
\includegraphics[width=0.9\textwidth]{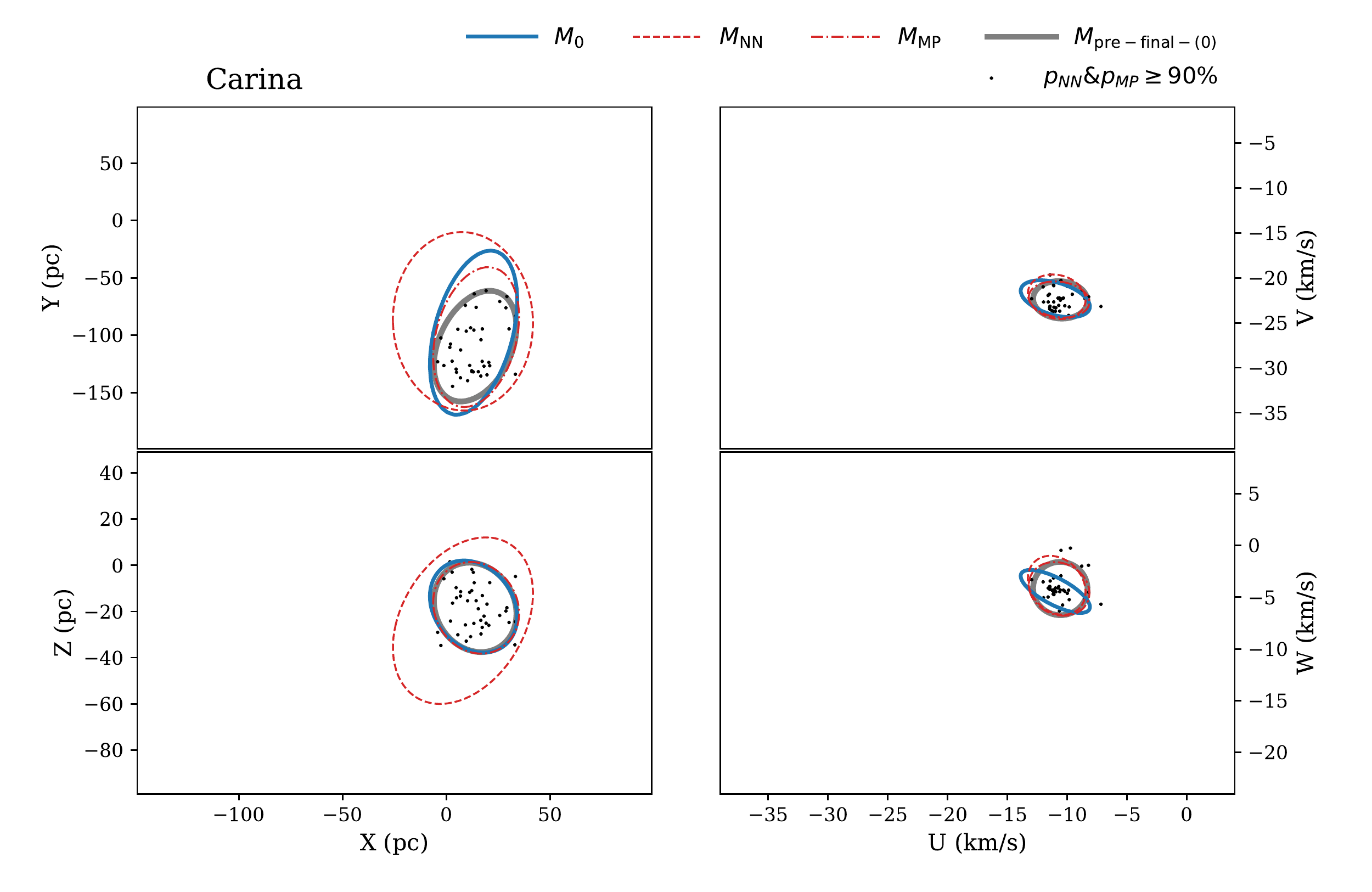}
\caption{Same to Fig.~\ref{fig:twa} but for Carina.}
\label{fig:carina}
\end{figure*}

\begin{table}
\scriptsize
\setlength\tabcolsep{3pt} 
\begin{threeparttable}
\begin{tabular}{*{6}{c}}
\hline
Group & Name & SpT& \ra &\dec  &  p\tnote{a} \\
& & & hh:mm:ss & dd:mm:ss & per cent  \\
 \hline \hline
 \multicolumn{6}{c}{\it bona fide members} \\
    Carina &                     CD-57 1709 &         K0V &  07:21:23.71 &  -57:20:37.0 &  99.2 \\ 
    Carina &                     CD-42 3328 &        K1IV &  07:33:21.17 &  -42:55:42.3 &  92.2 \\ 
    Carina &                      CD-63 336 &          G2 &  07:42:38.90 &  -63:36:14.0 & 100.0 \\
\multicolumn{6}{c}{\dots} \\
\multicolumn{6}{c}{\dots} \\
\multicolumn{6}{c}{\dots} \\
%\hline
 \multicolumn{6}{c}{\it probable members} \\
    Carina &                     J0715-6555 &          M5 &  07:15:17.05 &  -65:55:48.7 &  93.1 \\ 
    Carina &        2MASS J08035086-7344331 &        M3.7 &  08:03:50.85 &  -73:44:33.1 &  99.8 \\ 
    Carina &        2MASS J08063045-6809212 &        M3.8 &  08:06:30.46 &  -68:09:21.2 & 100.0 \\ 
\multicolumn{6}{c}{\dots} \\
\multicolumn{6}{c}{\dots} \\
\multicolumn{6}{c}{\dots} \\
\hline
\end{tabular}
\begin{tablenotes}
\item[a] Membership probability using \mfin.
\end{tablenotes}
\end{threeparttable}
\caption{Selected Carina members in 2 membership classes.
The description of classes are same as Table~\ref{tab:memb_bpmg}.
The entire table is available online. }
\label{tab:memb_carina}
\end{table}

In the entire data set, 191 stars have been claimed as Carina members.
There are 37 {\it bona fide} and 31 {\it probable members}.
These 2 sets of members are listed in Table~\ref{tab:memb_carina}.

On the other hand, 17 stars appear to be not associated with any NYMGs ($p_{\rm field}>$ 80 per cent).
Sixty stars likely belong to other NYMGs ($p_{\rm other group}>$ 80 per cent), and 12 of these 60 stars have already suggested as members of other NYMGs.
Forty-eight stars claimed as Carina members in the literature are claimed as members of other NYMGs for the first time in this study.

\subsection{Columba}

\begin{figure*}
\includegraphics[width=0.9\textwidth]{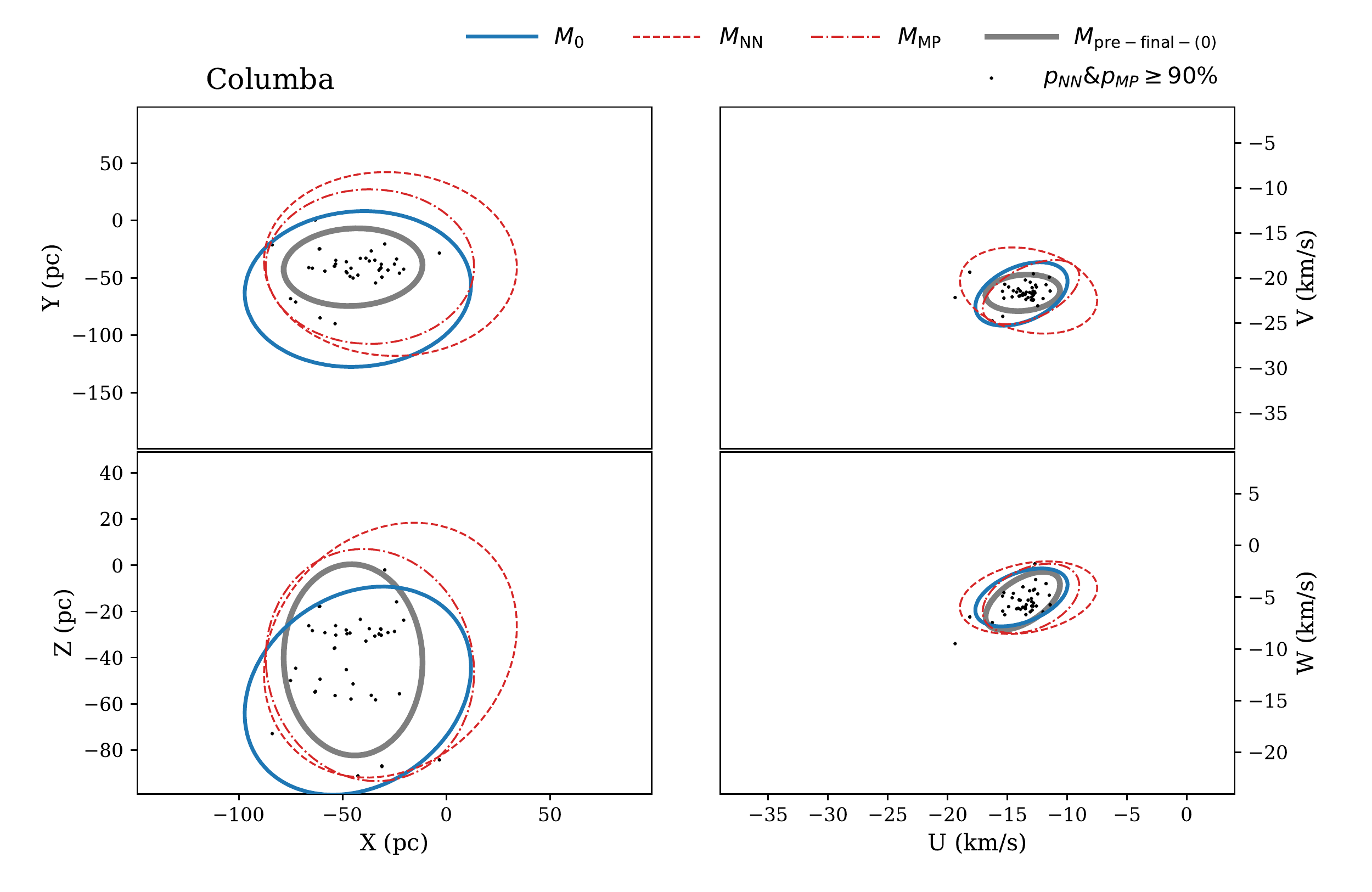}
\caption{Same to Fig.~\ref{fig:twa} but for Columba}
\label{fig:columba}
\end{figure*}

\begin{table}
\scriptsize
\setlength\tabcolsep{3pt} 
\begin{threeparttable}
\begin{tabular}{*{6}{c}}
\hline
Group & Name & SpT& \ra &\dec  &  p\tnote{a} \\
& & & hh:mm:ss & dd:mm:ss & per cent  \\
 \hline \hline
 \multicolumn{6}{c}{\it bona fide members} \\
   Columba &                         HD8077 &         F6V &  01:19:05.61 &  -53:51:01.9 &  90.8 \\ 
   Columba &                      CD-52 381 &        K2Ve &  01:52:14.63 &  -52:19:33.2 &  98.9\\ 
   Columba &        2MASS J01540267-4040440 &         K7V &  01:54:02.69 &  -40:40:44.2 &  93.9 \\ 
\multicolumn{6}{c}{\dots} \\
\multicolumn{6}{c}{\dots} \\
\multicolumn{6}{c}{\dots} \\
%\hline
 \multicolumn{6}{c}{\it highly likely members} \\
   Columba &        2MASS J02303239-4342232 &        K5Ve &  02:30:32.42 &  -43:42:23.4 &  82.1 \\
   Columba &        WASP 050206.19+311102.2 &          K2 &  05:02:06.19 &  +31:11:02.4 &  80.0 \\ 
   Columba &        2MASS J05432676-3025129 &       M0.5V &  05:43:26.76 &  -30:25:13.0 &  88.4\\ 
 \multicolumn{6}{c}{\it probable members} \\
   Columba &       WISE J032144.76-330949.5 &        M5.8 &  03:21:44.79 &  -33:09:48.9 &  98.4 \\ 
   Columba &       WISE J032443.06-273323.1 &        M5.5 &  03:24:43.05 &  -27:33:23.0 &  92.7\\ 
   Columba &                     J0326-3850 &          M3 &  03:26:37.04 &  -38:50:15.9 &  99.7 \\
\multicolumn{6}{c}{\dots} \\
\multicolumn{6}{c}{\dots} \\
\multicolumn{6}{c}{\dots} \\
 \multicolumn{6}{c}{\it possible members} \\
   Columba &                       HD 30447 &         F3V &  04:46:49.53 &  -26:18:08.9 & 100.0 \\ 
   Columba &                       HD 35841 &         F3V &  05:26:36.59 &  -22:29:23.7 & 100.0 \\
   Columba &        2MASS J05361998-1920396 &   L2{gamma} &  05:36:20.02 &  -19:20:40.1 &  99.4 \\ 
\hline
\end{tabular}
\begin{tablenotes}
\item[a] Membership probability using \mfin.
\end{tablenotes}
\end{threeparttable}
\caption{Selected Columba members in 4 membership classes. 
The description of classes are same as Table~\ref{tab:memb_bpmg}. 
The entire table is available online.}
\label{tab:memb_columba}
\end{table}

In the entire data set, 270 stars have been claimed as Columba members.
There are 87 {\it bona fide}, 3 {\it highly likely}, 64 {\it probable}, and 3 {\it possible members}.
These 4 sets of members are listed in Table~\ref{tab:memb_columba}.

On the other hand, 22 stars appear to be not associated with any NYMGs ($p_{\rm field}>$ 80 per cent).
Forty-one stars likely belong to other NYMGs ($p_{\rm other group}>$ 80 per cent), and 19 of these 41 stars have already suggested as members of other NYMGs.
Twenty-two stars claimed as Columba members in the literature are claimed as members of other NYMGs for the first time in this study.

\subsection{Argus}

\begin{figure*}
\includegraphics[width=0.9\textwidth]{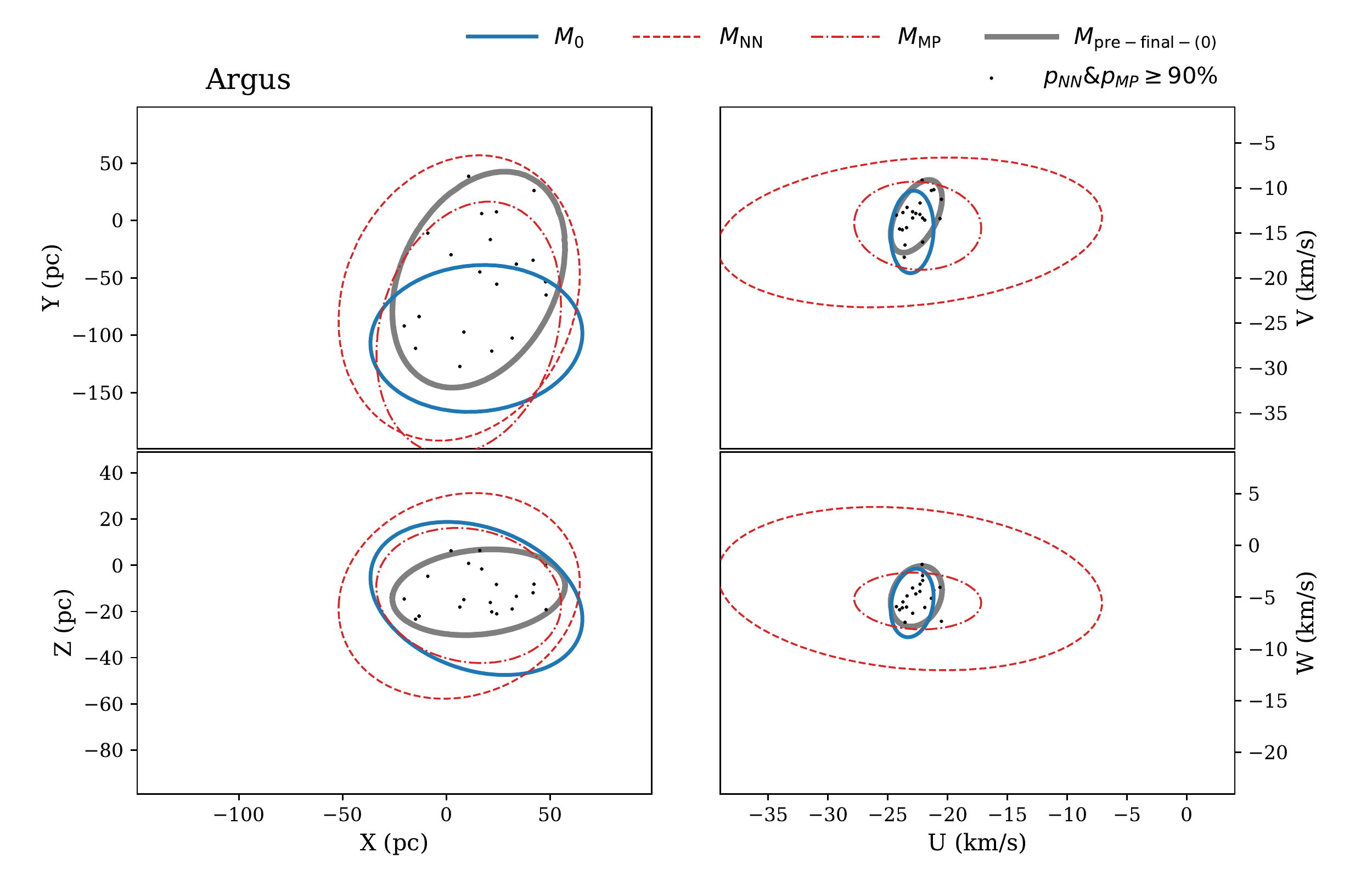}
\caption{Same to Fig.~\ref{fig:twa} but for Argus.}
\label{fig:argus}
\end{figure*}

\begin{table}
\scriptsize
\setlength\tabcolsep{3pt} 
\begin{threeparttable}
\begin{tabular}{*{6}{c}}
\hline
Group & Name & SpT& \ra &\dec  &  p\tnote{a} \\
& & & hh:mm:ss & dd:mm:ss & per cent  \\
 \hline \hline
 \multicolumn{6}{c}{\it bona fide members} \\
     Argus &                         FT Psc &   M3.5+M4.5 &  00:50:33.25 &  +24:49:00.2 &  94.7\\ 
     Argus &        2MASS J05090356-4209199 &       M3.5V &  05:09:03.56 &  -42:09:19.8 &  97.3 \\ 
     Argus &                     CD-29 2360 &        K3Ve &  05:34:59.23 &  -29:54:04.0 &  94.8\\ 
\multicolumn{6}{c}{\dots} \\
\multicolumn{6}{c}{\dots} \\
\multicolumn{6}{c}{\dots} \\
 \multicolumn{6}{c}{\it highly likely members} \\
     Argus &                  TYC 8561-0970 &         K0V &  07:53:55.48 &  -57:10:07.3 &  84.6 \\ 
     Argus &                       PMM 2456 &         K3e &  08:35:43.69 &  -53:21:20.3 &  89.2 \\ 
     Argus &                      CD-74 673 &        K3Ve &  12:20:34.40 &  -75:39:28.8 &  87.7 \\ 
 \multicolumn{6}{c}{\it probable members} \\
     Argus &        2MASS J07140101-1945332 &       M4.5V &  07:14:01.01 &  -19:45:33.2 &  92.1 \\ 
     Argus &                       PMM 5314 &           A &  08:28:34.53 &  -52:37:03.5 & 100.0 \\ 
  \multicolumn{6}{c}{\it possible members} \\
    Argus &        2MASS J00452143+1634446 &    L2{beta} &  00:45:21.42 &  +16:34:44.7 &  99.7\\ 
     Argus &                         AP Col &          M5 &  06:04:52.15 &  -34:33:35.8 & 100.0 \\ 
     Argus &                       PMM 2888 &          F5 &  08:43:52.30 &  -53:13:59.8 &  98.6 \\
\multicolumn{6}{c}{\dots} \\
\multicolumn{6}{c}{\dots} \\
\multicolumn{6}{c}{\dots} \\
\hline
\end{tabular}
\begin{tablenotes}
\item[a] Membership probability using  \mfin.
\end{tablenotes}
\end{threeparttable}
\caption{Selected Argus members in 4 membership classes. 
The description of classes are same as Table~\ref{tab:memb_bpmg}.
The entire table is available online.}
\label{tab:memb_argus}
\end{table}

In the entire data set, 167 stars have been claimed as Argus members.
There are 46 {\it bona fide}, 3 {\it highly likely}, 2 {\it probable}, and 5 {\it possible members}.
These 4 sets of members are listed in Table~\ref{tab:memb_argus}.

On the other hand, 60 stars appear to be not associated with any NYMGs ($p_{\rm field}>$ 80 per cent).
Five stars likely belong to other NYMGs ($p_{\rm other group}>$ 80 per cent), and one of these 5 stars has already suggested as members of other NYMGs.
Four stars claimed as Argus members in the literature are claimed as members of other NYMGs for the first time in this study.

\subsection{ABDor}

\begin{figure*}
\includegraphics[width=0.9\textwidth]{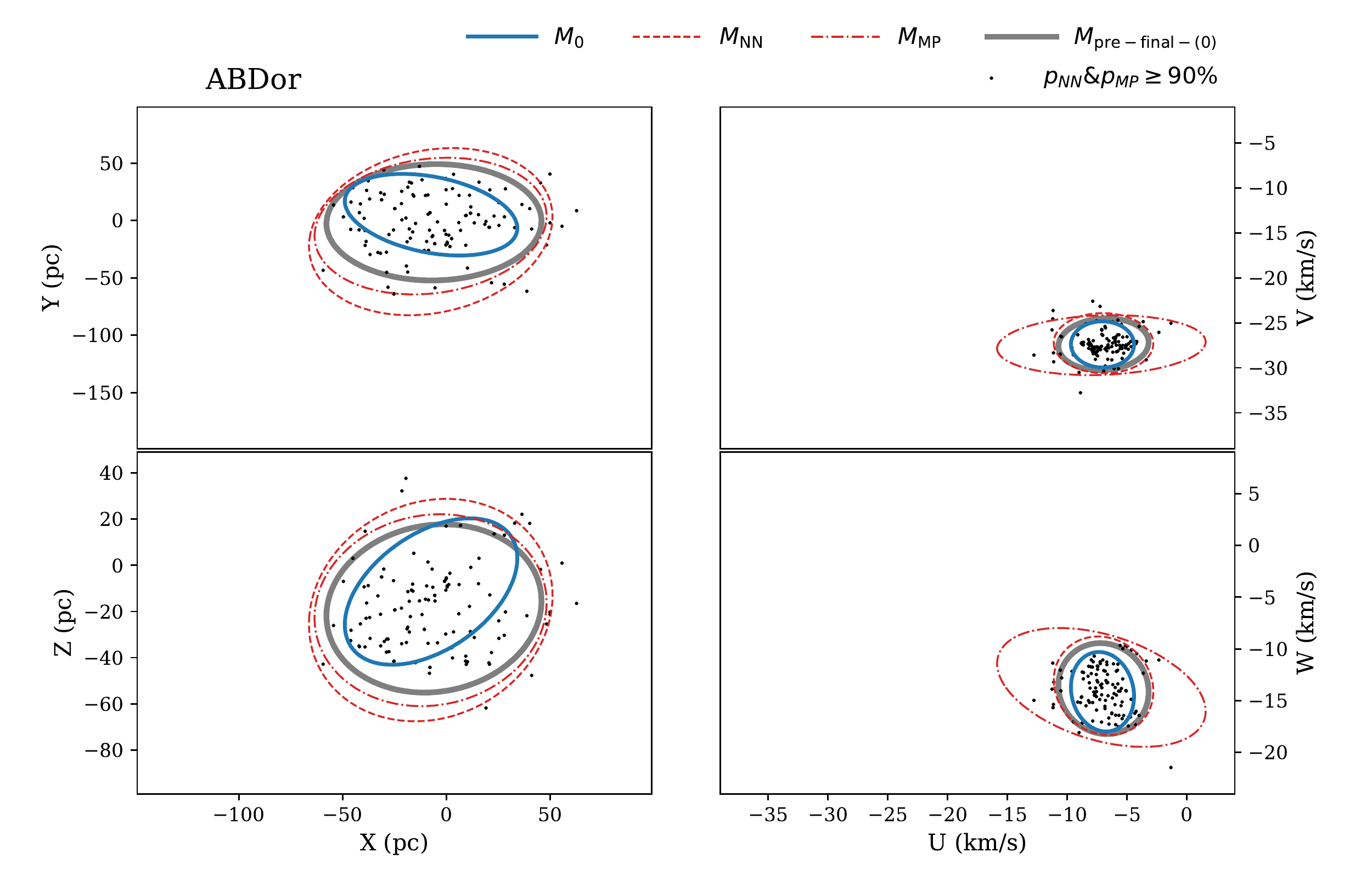}
\caption{Same to Fig.~\ref{fig:twa} but for ABDor.}
\label{fig:abdor}
\end{figure*}

\begin{table}
\scriptsize
\setlength\tabcolsep{3pt} 
\begin{threeparttable}
\begin{tabular}{*{6}{c}}
\hline
Group & Name & SpT& \ra &\dec  &  p\tnote{a} \\
& & & hh:mm:ss & dd:mm:ss & per cent  \\
 \hline \hline
 \multicolumn{6}{c}{\it bona fide members} \\
     ABDor &                         PW And &        K0Ve &  00:18:20.89 &  +30:57:22.1 & 100.0 \\ 
     ABDor &        2MASS J00192626+4614078 &          M8 &  00:19:26.27 &  +46:14:07.8 & 100.0 \\ 
     ABDor &                     J0025-0957 &       M3.0V &  00:25:50.98 &  -09:57:39.9 &  94.0\\
\multicolumn{6}{c}{\dots} \\
\multicolumn{6}{c}{\dots} \\
\multicolumn{6}{c}{\dots} \\
 \multicolumn{6}{c}{\it highly likely members} \\
     ABDor &                     J0139+2611 &          M4 &  01:39:58.71 &  +26:11:03.1 &  81.4 \\ 
     ABDor &                       G 133-40 &          M0 &  01:47:29.55 &  +34:13:09.2 &  80.7 \\ 
     ABDor &        2MASS J02105538-4603588 &        K4.2 &  02:10:55.39 &  -46:03:58.7 &  86.3 \\
          ABDor &                     CD-35 2749 &        K1Ve &  06:11:55.72 &  -35:29:12.9 &  89.9\\
     ABDor &                      HD 199058 &         G5V &  20:54:21.09 &  +09:02:23.8 &  89.4 \\ 
 \multicolumn{6}{c}{\it probable members} \\
     ABDor &                     J0014+4758 &          M6 &  00:14:18.23 &  +47:58:07.6 &  99.9 \\ 
     ABDor &                     J0021+4304 &          M5 &  00:21:46.97 &  +43:04:21.4 &  99.9 \\ 
     ABDor &                     J0024-2522 &          M3 &  00:24:32.02 &  -25:22:53.0 & 100.0 \\ 
\multicolumn{6}{c}{\dots} \\
\multicolumn{6}{c}{\dots} \\
\multicolumn{6}{c}{\dots} \\
 \multicolumn{6}{c}{\it possible members} \\
     ABDor &                        HIP4967 &           M &  01:03:40.14 &  +40:51:29.2 & 100.0 \\
     ABDor &                       GJ 2022A &        M4.0 &  01:24:27.64 &  -33:55:09.4 & 100.0 \\ 
     ABDor &                G 269-153 C (E) &        M5.0 &  01:24:30.42 &  -33:54:59.5 & 100.0 \\ 
     \multicolumn{6}{c}{\dots} \\
\multicolumn{6}{c}{\dots} \\
\multicolumn{6}{c}{\dots} \\
\hline
\end{tabular}
\begin{tablenotes}
\item[a] Membership probability using \mfin.
\end{tablenotes}
\end{threeparttable}
\caption{Selected ABDor members in 4 membership classes. 
The description of classes are same as Table~\ref{tab:memb_bpmg}. 
The entire table is available online.}
\label{tab:memb_abdor}
\end{table}

In the entire data set, 465 stars have been claimed as ABDor members.
There are 129 {\it bona fide}, 5 {\it highly likely}, 126 {\it probable}, and 20 {\it possible members}.
These 4 sets of members are listed in Table~\ref{tab:memb_abdor}.

On the other hand, 75 stars appear to be not associated with any NYMGs ($p_{\rm field}>$ 80 per cent).
Eighteen stars likely belong to other NYMGs ($p_{\rm other group}>$ 80 per cent), and two of these 18 stars have already suggested as members of other NYMGs.
Sixteen stars claimed as ABDor members in the literature are claimed as members of other NYMGs for the first time in this study.

\subsection{VCA}

\begin{figure*}
\includegraphics[width=0.9\textwidth]{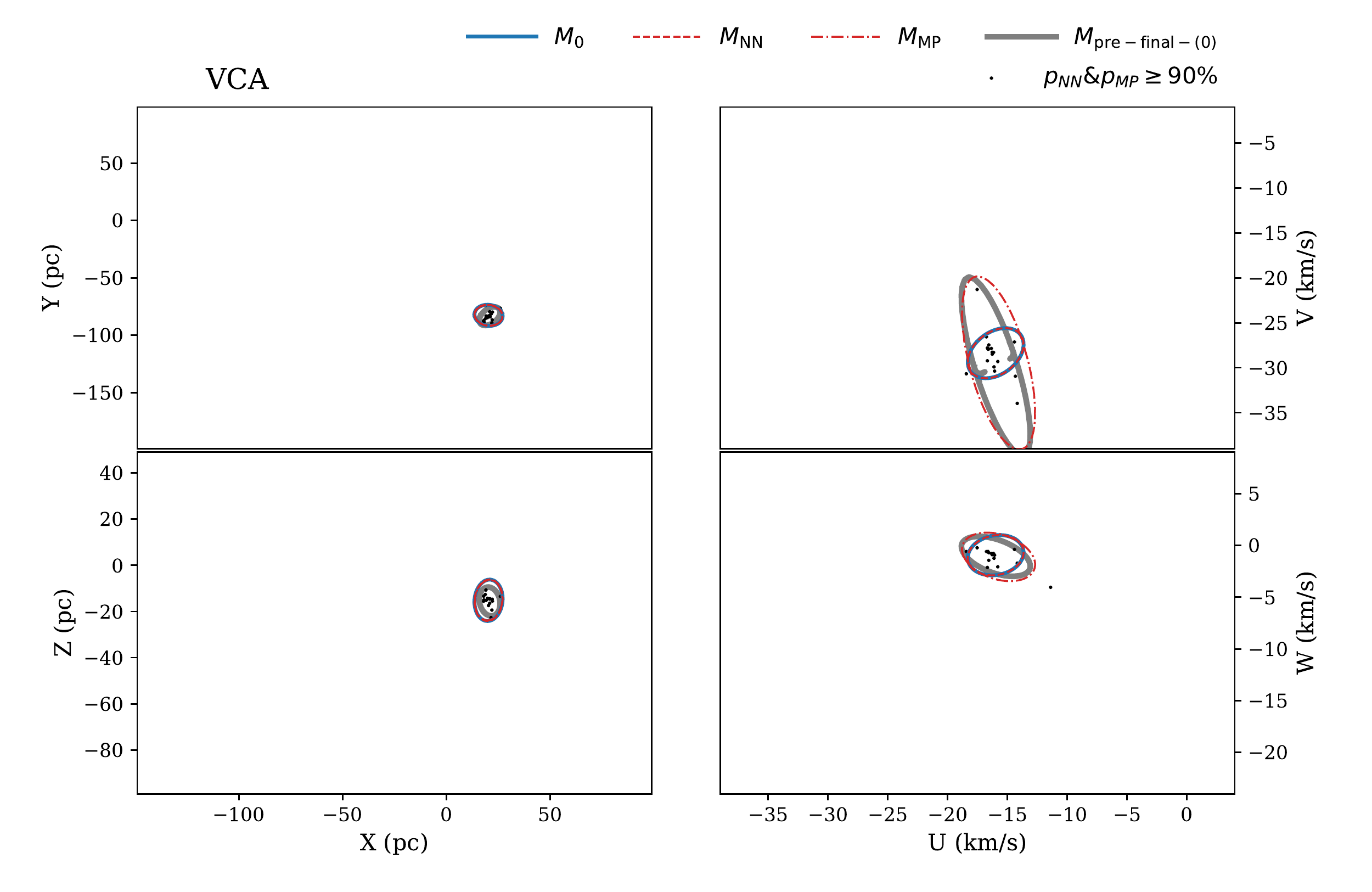}
\caption{Same to Fig.~\ref{fig:twa} but for VCA.}
\label{fig:vca}
\end{figure*}

\begin{table}
\scriptsize
\setlength\tabcolsep{3pt} 
\begin{threeparttable}
\begin{tabular}{*{6}{c}}
\hline
Group & Name & SpT& \ra &\dec  &  p\tnote{a} \\
& & & hh:mm:ss & dd:mm:ss & per cent  \\
 \hline \hline
 \multicolumn{6}{c}{\it bona fide members} \\
       VCA &                       HD 80563 &         F3V &  09:17:34.14 &  -63:23:14.5 & 100.0 \\ 
       VCA &                       HD 83946 &         F5V &  09:38:54.10 &  -64:59:26.7 & 100.0 \\ 
       VCA &                      HD 309681 &          G0 &  09:25:01.63 &  -64:37:31.2 & 100.0 \\ 
\multicolumn{6}{c}{\dots} \\
\multicolumn{6}{c}{\dots} \\
\multicolumn{6}{c}{\dots} \\
 \multicolumn{6}{c}{\it probable members} \\
       VCA &                 c Car (HR3571) &        B8II &  08:55:02.85 &  -60:38:40.7 & 100.0\\ 
       VCA &        2MASS J08383351-6716368 &          M5 &  08:38:33.51 &  -67:16:36.8 & 100.0 \\ 
       VCA &        2MASS J08485563-6113261 &          M5 &  08:48:55.63 &  -61:13:26.1 & 100.0 \\
\multicolumn{6}{c}{\dots} \\
\multicolumn{6}{c}{\dots} \\
\multicolumn{6}{c}{\dots} \\
\hline
\end{tabular}
\begin{tablenotes}
\item[a] Membership probability using \mfin.
\end{tablenotes}
\end{threeparttable}
\caption{Selected VCA members in 2 membership classes. 
The description of classes are same as Table~\ref{tab:memb_bpmg}.
The entire table is available online.}
\label{tab:memb_vca}
\end{table}

In the entire data set, 65 stars have been claimed as VCA members.
There are 19 {\it bona fide} and 42 {\it probable members}.
These 2 sets of members are listed in Table~\ref{tab:memb_vca}.

\subsection{Overall tendency}

For major NYMGs considered in our study including BPMG, among 1913 claimed members in the literature, $\sim$35 per cent of claimed members are retained as bona fide members.
About 13 per cent of members shown to be belonging to different groups and $\sim$14 per cent are estimated to be field star interlopers.

%%%%%%%%%%%%%%%%%%TABLES%%%%%%%%%%%%%%%%%%%%%%%%%%%%%%%%

% Don't change these lines
\bsp	% typesetting comment
\label{lastpage}
\end{document}